%% file: main.tex
\documentclass[reprint,
superscriptaddress,
nofootinbib,
amsmath,amssymb,
aps,
floatfix
]{revtex4-2}

\usepackage{graphicx}
\usepackage{dcolumn}
\usepackage{bm}
\usepackage{float}
\usepackage[switch*]{lineno}
\usepackage{xcolor}
\usepackage{adjustbox}
\usepackage{multirow}
\usepackage{physics}
\usepackage{mathtools}
\usepackage{tikz-feynman}
\usepackage{tikz}
\usetikzlibrary{decorations.pathmorphing,shapes}
\usepackage[bottom]{footmisc}

\def\gsim{\mathrel {\rlap{\lower4pt\hbox{\hskip1pt$\sim$}}
    \raise1pt\hbox{$>$}}}         

\graphicspath{{figures/}}

\begin{document}

\preprint{APS/123-QED}

\title{Simultaneous Inference of Effective Range Parameters and EFT Truncation Uncertainty in $^{3}$He-$\alpha$ Scattering}

\author{Andrius Burnelis}
	\email{ab351021@ohio.edu}
    \affiliation{Department of Physics and Astronomy and Institute of Nuclear and Particle Physics, Ohio University, Athens, Ohio 45701, USA}

\author{Daniel R. Phillips}
    \email{phillid1@ohio.edu}
    \affiliation{Department of Physics and Astronomy and Institute of Nuclear and Particle Physics, Ohio University, Athens, Ohio 45701, USA}
    \affiliation{Department of Physics, Chalmers University of Technology, SE-41296 G\"oteborg, Sweden}

\begin{abstract}
    We extend previous halo effective field theory analyses of low-energy elastic scattering of $^{3}$He-$^{4}$He, including the $\frac{7}{2}^{-}$ $f$-wave resonance as an explicit degree of freedom. The presence of this resonance necessitates a changing power counting scheme depending on the kinematic region. Therefore, we construct a theory uncertainty model at the partial wave amplitude level, allowing us to generate a sophisticated theory covariance matrix that captures the way the theory error structure changes as energy increases. We then perform a Bayesian analysis and simultaneously estimate the joint posterior distributions of the effective range theory parameters and the parameters that characterize the effective field theory truncation uncertainty. We compare two different analyses: no $f$-wave interactions for data up to $E_{\text{max}} = 2.6$ MeV, and including $f$-wave interactions for data up to $E_{\text{max}} = 5.5$ MeV. The inferred breakdown scales in each analysis are consistent with previous work. We find that $f$-wave interactions are needed to describe data for $E_{lab} \gsim 3.6$ MeV.
\end{abstract}

\maketitle

\input{introduction.tex}

\input{effective_field_theory.tex}

\input{data.tex}

\input{scattering_model.tex}

\input{uncertainty_model.tex}

\input{simultaneous_sampling.tex}

\input{results_no_f_waves.tex}

\input{results_f_waves.tex}

\input{conclusion.tex}

\section{Acknowledgements}
We thank Alexandra Semposki,  Maheshwor Poudel, and Som Paneru for helpful discussions. This work was supported by the U.S. Department of Energy under contract DE-FG02-93ER40756 and by the National Science Foundation CSSI program under award number OAC-2004601 (BAND Collaboration). 

\section{Data Availability}
The data and code used in this work are available at \url{https://github.com/AndriusBurn/3he_alpha_SPF}.

\appendix{\input{width.tex}}

\bibliographystyle{unsrt}
\bibliography{citations}

\end{document}

%% file: introduction.tex
\section{Introduction}
\label{sec:intro}
Low-energy elastic scattering of light nuclei is important since measurements of low-energy scattering cross sections provide valuable information about nuclear interactions and reaction rates. These reactions are responsible for many stellar processes such as nucleosynthesis and neutrino production.

Understanding the low-energy elastic scattering of $^{3}$He and $^{4}$He is crucial for understanding the solar $pp$-II and $pp$-III chains. The $^{3}$He($\alpha$, $\gamma$)$^{7}$Be reaction is responsible for the production of $^{7}$Be, which then either undergoes proton or electron capture. Both reactions contribute to the solar neutrino flux~\cite{Adelberger2011}. The reaction rate of this process may be determined from the astrophysical $S$-factor. The $S$-factor is a scaled total cross section which is dependent on energy, and is then extrapolated to energies relevant to stellar environments or the Gamow peak~\cite{Adelberger2011, Paneru,Wiescher:2025}.

The composite $^{7}$Be nucleus exhibits $^{3}$He and $^{4}$He clustering. The energy scales of the $^{3}$He and $^{4}$He constituents are much higher than the energy scales associated with the $^{7}$Be nucleus. This separation of scales makes the $^{3}$He-$^{4}$He system an ideal candidate for analysis with halo effective field theory (EFT)~\cite{Higa:2018, Poudel_2022}, since amplitudes can be expanded in the ratio $Q = \frac{M_{lo}}{\Lambda_{B}}$ where $M_{lo}$ and $\Lambda_{B}$ are the low- and high-energy scales, respectively.
 Halo EFT is thus a powerful tool that provides a rigorous and systematic approach to halo/clustered systems, allowing for controlled approximations and uncertainty quantification~\cite{Hammer:2017}. The halo EFT framework allows for the organization of terms in order of decreasing importance, providing a built-in way to quantify uncertainties. Here we calibrate the EFT model uncertainty using the convergence pattern of the EFT and the model discrepancy between the EFT and the data.

Recent data for low-energy elastic scattering of $^{3}$He and $^{4}$He was obtained by Paneru \textit{et al.}~\cite{Paneru} using the Scattering of Nuclei in Inverse Kinematics (SONIK) scattering chamber at TRIUMF. This dataset consists of differential cross section measurements with $^{3}$He (lab) beam energies ranging from 0.7 to 5.5 MeV, impinged on a $^{4}$He gas target. This data extends previous measurements of this system~\cite{Barnard:1964, Spiger_1967, Boykin:1972} to lower energies with greater precision.

Theoretical work on the elastic scattering and capture reactions of ${}^3$He and ${}^4$He nuclei has a long history~\cite{Christy:1961,Parker:1963}. Work prior to 2011 is reviewed in Ref.~\cite{Adelberger:2011}. Since then the $R$-matrix formalism has been used extensively to evaluate and extrapolate these two reactions, treating them as a coupled-channels problem (see, e.g., Refs.~\cite{deboer2014monte, Odell:2022}). The last ten years have also seen increasingly more sophisticated {\it ab initio} treatments of these processes within the No Core Shell Model plus Continuum (NCSMC) framework~\cite{DohetEraly:2016,Atkinson:2025}. 

Higa \textit{et al.}~\cite{Higa:2018} were the first to apply halo effective field theory (EFT) to the low-energy radiative capture process $^{3}$He($\alpha, \gamma$)$^{7}$Be and extract the astrophysical $S$-factor. Subsequent EFT analyses were carried out in Refs.~\cite{Zhang:2020, Premarathna:2020,Khadka:2025hef}. However, none of these works calibrated the EFT directly to scattering cross sections. Instead, they focused on the capture reaction, although Refs.~\cite{Higa:2018,Premarathna:2020,Khadka:2025hef} used the phase shifts of Ref.~\cite{Boykin:1972} as input. 

In Ref.~\cite{Poudel_2022}, Poudel \& Phillips used Halo EFT to analyze data on ${}^3$He-${}^4$He elastic scattering. However, they included the effects of the $\frac{7}{2}^{-}$ resonance via the phenomenological $R$-matrix method. While the analysis of Ref.~\cite{Poudel_2022} is successful in describing the low-energy scattering data, combining Halo EFT with the phenomenological $R$-matrix formalism does not provide a clean uncertainty quantification. In this work we extend the analysis performed in~\cite{Poudel_2022} by directly including the $\frac{7}{2}^{-}$ resonance in our halo EFT. In parallel, we develop a sophisticated uncertainty model and then simultaneously sample the physics parameters alongside the EFT truncation uncertainty parameters.

In Sec.~\ref{sec:eft} we justify the use of halo effective field theory in the $^{3}$He-$^{4}$He system and review the framework and its application. We also develop the power counting scheme that is used to organize the contributions to the scattering amplitude. Then, in Sec.~\ref{sec:data}, we discuss the data used in this analysis. In Sec.~\ref{sec:scattering_model} we describe the scattering model in detail and list the different relevant partial wave channels. In Sec.~\ref{sec:uncertainty_model} we introduce the statistical model that is used to quantify the theory uncertainties, and construct the theory covariance matrix. Sec.~\ref{sec:simultaneous_sampling} derives the desired joint posterior distribution for the scattering model parameters and the effective field theory truncation uncertainty parameters. In Sec.~\ref{sec:implementation_without_f_waves} we present the implementation and results of an analysis that does not include the $f$-waves. We then outline the changes necessary to include the $f$-waves, implement the $f$-wave analysis, and present results in Sec.~\ref{sec:implementation_with_f_waves}. Finally, we provide a summary of our results and discuss potential future work in Sec.~\ref{sec:conclusion}. 

%% file: effective_field_theory.tex
\section{Effective Field Theory}
\label{sec:eft}

\subsection{Energy Scales}
Effective field theory (EFT) is a powerful framework that allows us to exploit a separation of scales in order to perform a systematic expansion in a dimensionless parameter $Q$, defined as $Q = \frac{M_{lo}}{\Lambda_{B}}$. 
Low energy $^{3}$He-$^{4}$He scattering is a candidate for EFT due to its appreciable separation of  scales. The scales that are relevant for this system are the energy scales associated with the low-lying states in the $^{7}$Be system and the energy scales and radii of its constituents (the $^{3}$He and $^{4}$He nuclei). 

\begin{figure}
    \centering
    \includegraphics[width = 0.9\columnwidth]{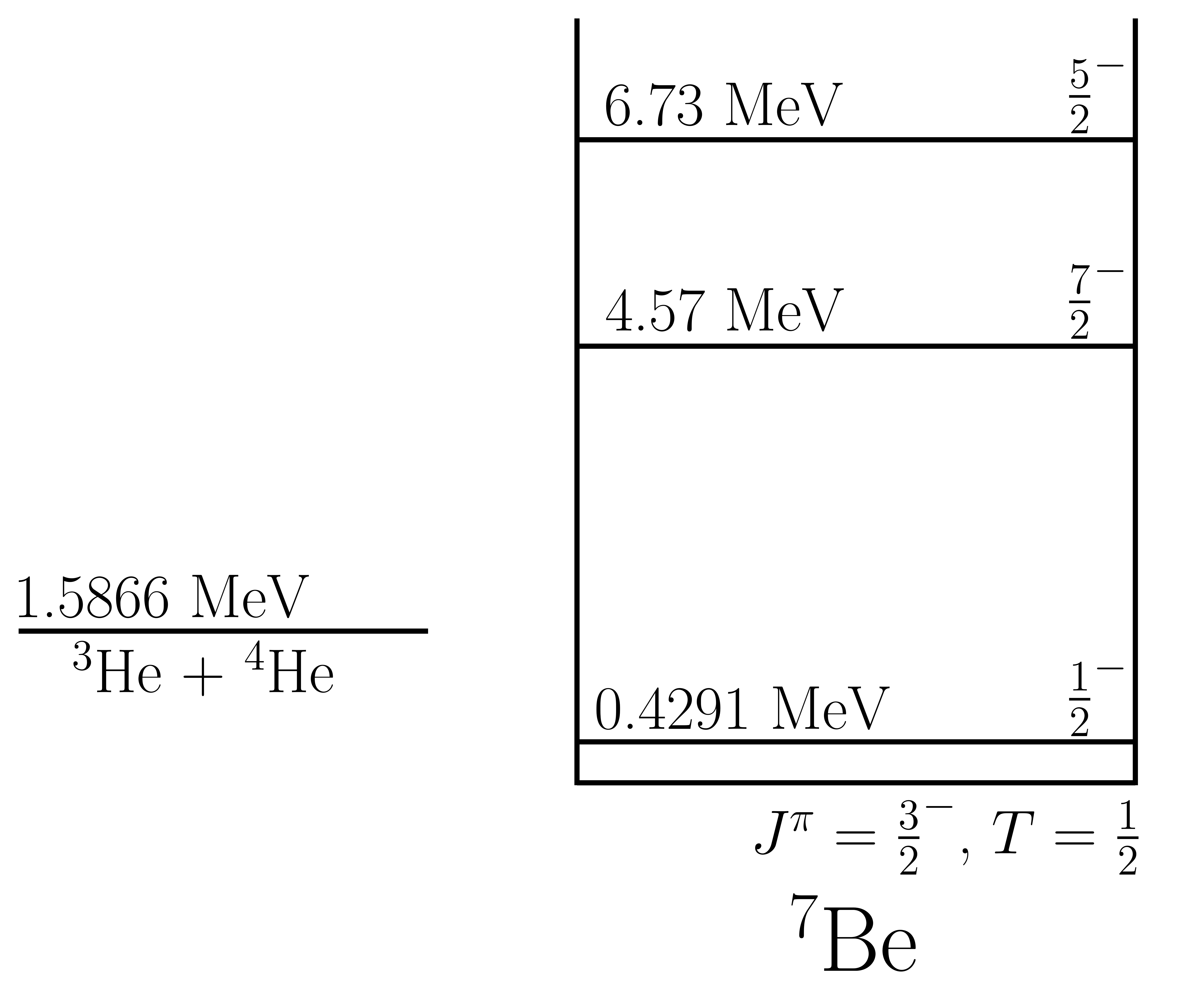}
    \caption{\label{fig:level_diagram}Energy level diagram of the $^{7}$Be system~\cite{Tilley2002}.}
\end{figure}

The energy level diagram of the $^{7}$Be system is shown in Fig.~\ref{fig:level_diagram}~\cite{Tilley2002}. We label the states according to their total angular momentum $j = \ell \pm \frac{1}{2}$, where $\ell$ is the orbital angular momentum (partial wave number) and the $\pm \frac{1}{2}$ denotes the coupling to the $^{3}$He spin.

There are two low-lying $p$-wave ($\ell = 1$) states in $^{7}$Be, the $\frac{3}{2}^{-}$ ground state at 1.5866 MeV below the scattering threshold and the $\frac{1}{2}^{-}$ bound state at 1.1575 MeV below the threshold~\cite{Tilley2002}. In what follows, the $\frac{3}{2}^{-}$ channel corresponds to the $+$ superscript or subscript, and the $\frac{1}{2}^{-}$ channel corresponds to the $-$ superscript or subscript. The next state in the $^{7}$Be system is the $\frac{7}{2}^{-}$ ($\ell = 3$) resonance at 2.983 MeV above the threshold~\cite{Tilley2002}. These states are the low energy scales. There is also a $\frac{5}{2}^{-}$ resonance at 5.143 MeV above the threshold~\cite{Tilley2002}. We refer to the $\frac{7}{2}^{-}$ channel with a $+$ superscript or subscript, and the $\frac{5}{2}^{-}$ channel with a $-$ superscript or subscript. 

Meanwhile, the high energy scales are associated with the constituents. The $^{3}$He nucleus has a binding energy of 7.72 MeV and a proton separation energy of 5.49 MeV~\cite{Purcell:2010}, as well as a radius of 1.97 fm~\cite{Angeli:2013epw}. The $^{4}$He nucleus has a large binding energy of 28.30 MeV and a radius of 1.68 fm~\cite{Angeli:2013epw}. The large separation of scales between the low energy scales in $^{7}$Be and the high energy scales associated with its constituents justifies the use of effective field theory to describe the low-energy scattering of $^{3}$He and $^{4}$He.

\subsection{Halo Effective Field Theory and Effective Range Theory}
We start with the EFT framework which allows us to systematically include the relevant degrees of freedom and then organize the terms in the scattering amplitude according to their relative importance. In this framework we treat the $^{3}$He and $^{4}$He nuclei as fundamental degrees of freedom. We then also include the low-lying states of $^{7}$Be as degrees of freedom. The $^{3}$He nucleus is represented as the $\psi$ field, and the $^{4}$He nucleus is represented as the $\phi$ field. The low-lying $^{7}$Be bound states will be represented as dimer fields, denoted $D_{\ell}^{\pm}$. We also introduce an $s$-wave dimer $D_0$, in order to easily accommodate the large $s$-wave scattering length between ${}^3$He and ${}^4$He nuclei in the EFT. 

As discussed in Ref.~\cite{Poudel_2022}, the Lagrangian for this system is composed of the free Lagrangian $\mathcal{L}_{\text{free}}$, the Coulomb Lagrangian $\mathcal{L}_{C}$, the $s$-wave interactions, $\mathcal{L}_{s}$, and the $p$-wave interactions $\mathcal{L}_{p^{\pm}}$. In this work we extend the EFT to include an $f$-wave Lagrangian $\mathcal{L}_{f^{\pm}}$. 

We write the free Lagrangian as
\begin{align}
	\label{eq:free_lagrangian}
	\mathcal{L}_{\text{free}} = \psi^{\dagger a} \Bigg( i \frac{\partial}{\partial t} + & \frac{\nabla^{2}}{2 m_{\psi}} \Bigg) \psi_{a} + \phi^{\dagger} \Bigg( i \frac{\partial}{\partial t} + \frac{\nabla^{2}}{2 m_{\phi}} \Bigg) \phi \nonumber \\
	+ \sum_{\ell = 0, 1, 3} (D_{\ell}^{\pm})^{\dagger b_{\ell}^{\pm}} & \Bigg( \omega_{\ell}^{\pm} \Big[ i \frac{\partial}{\partial t} + \frac{\nabla^{2}}{2 m} \Big] \nonumber \\
	+ \Xi_{\ell}^{\pm} \Big[ i \frac{\partial}{\partial t} + & \frac{\nabla^{2}}{2 m} \Big]^{2} + \cdots + \Delta_{\ell}^{\pm} \Bigg)(D_{\ell}^{\pm})_{b_{\ell}^{\pm}}.
\end{align}
We will sum over repeated indices. In this expression, $m_{\psi} = 2809.43$ MeV, $m_{\phi} = 3728.42$ MeV, $m = m_{\psi} + m_{\phi}$, $\omega_{\ell}^{\pm} = \pm 1$ depending on the sign of the effective range in that channel, $\Delta_{\ell}^{\pm}$ is the bare dimer binding energy, and $\Xi_{\ell}^{\pm}$ is a low energy constant with dimensions of inverse energy. The ellipses represent operators that produce higher-order terms in the expansion. We have also included the indices $a$ and $b_{\ell}^{\pm}$ to indicate the spinor indices of the $^{3}$He and dimer fields respectively. Note that the $^{\pm}$ notation has been slightly abused in Eq.~(\ref{eq:free_lagrangian}), as there are no $^+$ and $^-$ components for $s$-waves, and so only one dimer field in that channel. 

The $s$-wave interaction Lagrangian is given by
\begin{equation}
	\label{eq:s_wave_lagrangian}
	\mathcal{L}_{s} = g_{0} (D_{0}^{\dagger b_{0}} \psi_{b_{0}} \phi + \text{h.c.})
\end{equation}
where $g_{0}$ is the coupling strength of the $s$-wave interaction. The $p$-wave Lagrangian is given by
\begin{align}
    \label{eq:p_wave_lagrangian}
    \mathcal{L}_{p^{\pm}} = g_{1}^{\pm} \sum_{b_{1}^{\pm}, \lambda, \omega} \bigg\langle \frac{1}{2} \lambda; 1 \omega \bigg| 1^{\pm} b_{1}^{\pm} \bigg\rangle \bigg[ \bigg( \frac{2 m_{\psi}}{m} \psi_{\lambda} (i \nabla_{\omega} \phi) \nonumber \\ 
    - \frac{2 m_{\phi}}{m} (i \nabla_{\omega} \psi_{\lambda}) \phi \bigg) (D_{1}^{\pm})^{\dagger b_{1}^{\pm}} + \text{h.c.} \bigg]
\end{align}
where $g_{1}^{\pm}$ is the coupling strength of the $p$-wave interaction, and $\langle \frac{1}{2} \lambda; 1 \omega | 1^{\pm} b_{1}^{\pm} \rangle$ are the Clebsch-Gordan coefficients that couple the spin and angular momentum~\cite{Poudel_2022}. The notation $1^+$ ($1^-$) denotes the total angular momentum of 3/2 (1/2) in the two different $p$-wave channels indicated by the superscript $\pm$.

The $f$-wave Lagrangian now involves three derivatives acting on the fields. We define the Galilean invariant three-derivative operator $O_{\omega_{1}, \omega_{2}, \omega_{3}; \lambda}$ as
\begin{align}
    & O_{\omega_{1}, \omega_{2}, \omega_{3}; \lambda} = -i {\left( \frac{2 m_{\psi}}{m} \right)}^{3} \psi_{\lambda} (\nabla_{\omega_{1}} \nabla_{\omega_{2}} \nabla_{\omega_{3}} \phi) \nonumber \\
    & + i {\left( \frac{2 m_{\psi}}{m}\right)}^{2} \left( \frac{2 m_{\phi}}{m}\right) \left[ (\nabla_{\omega_{1}} \psi_{\lambda}) (\nabla_{\omega_{2}} \nabla_{\omega_{3}} \phi) + \text{perm.} \right] \nonumber \\
    & -i \left( \frac{2 m_{\psi}}{m}\right) {\left( \frac{2 m_{\phi}}{m}\right)}^{2} \left[ (\nabla_{\omega_{1}} \nabla_{\omega_{2}} \psi_{\lambda}) (\nabla_{\omega_{3}} \phi) + \text{perm.} \right] \nonumber \\
    & + i {\left( \frac{2 m_{\phi}}{m}\right)}^{3} (\nabla_{\omega_{1}} \nabla_{\omega_{2}} \nabla_{\omega_{3}} \psi_{\lambda}) \phi.
\end{align}
In this expression, ``perm.'' indicates that we include all permutations of the indices $\omega_{1}, \omega_{2}$, and  $\omega_{3}$. This operator has both rank 3 and rank 1 components. We then define the symmetric, rank 3 projection of this derivative operator as
\begin{equation}
    G_{\Omega; \lambda} = O_{[\omega_{1}, \omega_{2}, \omega_{3}] 3 \Omega; \lambda}
\end{equation}
and thus the $f$-wave Lagrangian is given by
\begin{align}
    \label{eq:f_wave_lagrangian}
    \mathcal{L}_{f^{\pm}} = g_{3}^{\pm} \sum_{b_{3}^{\pm}, \lambda, \Omega} \bigg\langle \frac{1}{2} \lambda; & 3 \Omega \bigg| 3^{\pm} b_{3}^{\pm} \bigg\rangle \nonumber \\
    \times & \bigg[ G_{\Omega; \lambda} (D_{3}^{\pm})^{\dagger b_{3}^{\pm}} + \text{h.c.} \bigg].
\end{align}
Here, $g_{3}^{\pm}$ is the coupling strength of the $f$-wave interaction, and $\langle \frac{1}{2} \lambda; 3 \Omega | 3^{\pm} b_{3}^{\pm} \rangle$ are the Clebsch-Gordan coefficients that couple the spin and angular momentum. Again, the notation $3^{\pm}$ indicates the total angular momentum of the two different $f$-wave channels: $3^+=7/2$, $3^-=5/2$. 

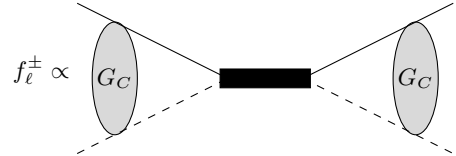
\begin{figure}[h]
    \begin{align}
        \centering
        f_{\ell}^{\pm} \propto
        \begin{tikzpicture}[baseline]
            \begin{feynman}
                \vertex (a) at (0, 1);
                \vertex (b) at (0, -1);
                \draw [fill = gray!30] (0.5, 0) ellipse (0.3 cm and 0.75 cm);
                \vertex (c) at (2, 0);
                \filldraw [draw = black] (1.9, -0.125) rectangle (3.1, 0.125);
                \vertex (d) at (3, 0);
                \draw [fill = gray!30] (4.5, 0) ellipse (0.3 cm and 0.75 cm);
                \vertex (e) at (5, 1);
                \vertex (f) at (5, -1);
                \node at (0.5, 0) {$G_{C}$};
                \node at (4.5, 0) {$G_{C}$};
                \diagram* {
                    (a) -- [plain] (c),
                    (b) -- [dashed] (c);
                    (d) -- [plain] (e),
                    (d) -- [dashed] (f)
                    };
            \end{feynman}
        \end{tikzpicture}
        \nonumber
    \end{align}
    \caption{\label{fig:partial_wave_amp} Graphical representation of the partial wave scattering amplitude $f_{\ell}^{\pm}$. The dashed lines represent the $^{3}$He field $(\psi_{\lambda})$, while the solid lines represent the $^{4}$He field $(\phi)$, and the filled rectangle is the dressed dimer field propagator for that partial wave.}
\end{figure}

Thus, the full Lagrangian is then
\begin{equation}
    \mathcal{L} = \mathcal{L}_{\text{free}} + \mathcal{L}_{C} + \mathcal{L}_{s} + \mathcal{L}_{p^{\pm}} + \mathcal{L}_{f^{\pm}}.
\end{equation}
Since the Lagrangian is organized according to partial waves, it is natural to express the full scattering amplitude in terms of the partial wave amplitudes. Each partial wave amplitude can be expressed in terms of the corresponding phase shift $\delta_{\ell}^{\pm}$, 
\begin{equation}
    \label{eq:f_ell_pm}
    f_{\ell}^{\pm} \equiv \frac{1}{\cot \delta_{\ell}^{\pm} - i}.
\end{equation}
A graphical representation of the partial wave scattering amplitude is shown in Fig.~\ref{fig:partial_wave_amp}.

\begin{figure}[h]
    \begin{align}
        \centering
        \begin{tikzpicture}[baseline]
            \filldraw [draw = black] (0, 0) rectangle (1, 0.25);
        \end{tikzpicture}
        = 
        \begin{tikzpicture}[baseline]
            \draw [draw = black] (0, 0) rectangle (1, 0.25);
        \end{tikzpicture} 
        +
        \begin{tikzpicture}[baseline]
            \draw [draw = black] (0, 0) rectangle (1, 0.25);
            \fill[gray!30] (1.5, 0.125) ellipse (0.5 cm and 0.5 cm);
            \begin{feynman}
                \vertex (a) at (1, 0.125);
                \vertex (b) at (2, 0.125);
                \vertex (c) at (1.5, 0.125) {$\Sigma_{C}$};
                \diagram*{
                    (a) -- [plain, half left, looseness = 1.7] (b) -- [dashed, half left, looseness = 1.7] (a)};
            \end{feynman}
            \filldraw [draw = black] (2, 0) rectangle (3, 0.25);
        \end{tikzpicture}
        \nonumber
    \end{align}
    \caption{\label{fig:dressing} Graphical representation of the dressing of the bare dimer propagator. The dashed lines represent the $^{3}$He field $(\psi_{\lambda})$, while the solid lines represent the $^{4}$He field $(\phi)$. The thick solid line represents the dressed dimer propagator.}
\end{figure}
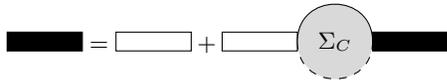

We dress the bare dimer propagators by summing the geometric series of bubble diagrams. The self-energy $\Sigma_{C}$ is calculated by summing the bubble diagrams with the Coulomb propagator, $G_C$. Once the dressed propagators are computed, we can then compute the scattering amplitude for each partial wave channel. A graphical representation of this dressing process is shown in Fig.~\ref{fig:dressing}. 

The scattering amplitude obtained from the EFT reproduces the Coulomb modified effective range expansion (CM-ERE)~\cite{Bertulani:2002, Bedaque:2003, Zhang:2017}. The CM-ERE allows us to parameterize the phase shifts in terms of the effective range parameters (ERPs). There is a mapping from the low energy constants that appear in the Lagrangian and these ERPs~\cite{Zhang:2020}. We have
\begin{align}
    \label{eq:invamp}
    k^{2 \ell + 1} ( \cot \delta_{\ell}^{\pm} - i) = \frac{2 k_{c}}{e^{-\pi \eta}} \Bigg[ \frac{\Gamma{(\ell + 1)}^{2} k_{c}^{2 \ell} K_{\ell}^{\pm}(k)}{(\ell^{2} + \eta^{2}) |\Gamma(\ell + i \eta)|^{2}} \nonumber \\ - \frac{k^{2 \ell} H(\eta)}{|\Gamma(1 + i\eta)|^{2}} \Bigg],
\end{align}
with 
\begin{eqnarray}
    H(\eta) = \Psi(i \eta) + \frac{1}{2 i \eta} - \ln(i \eta).
\end{eqnarray}
In these expressions, $\Psi$ is the digamma function and $\Gamma$ is the usual gamma function. The ERPs appear in the function $K_{\ell}^{\pm}(k)$, which is given by
\begin{flalign}
    K_{\ell}^{\pm} = \frac{1}{2 k_{c}^{2 \ell + 1}} \left[-\frac{1}{a_{\ell}^{\pm}} + \frac{r_{\ell}^{\pm}}{2} k^{2} + \frac{P_{\ell}^{\pm}}{4} k^{4} + \mathcal{O}(k^{6}) \right],&&
\end{flalign}
and is sometimes called the effective range function. We define $\eta = \frac{k_{c}}{k}$ with $k_{c} \equiv z_{1} z_{2} \alpha \mu$ where $z_{1}$ and $z_{2}$ are the charge numbers, $\alpha$ is the fine-structure constant, and $\mu$ is the reduced mass. The coefficients of the various powers of $k$ are the ERPs for the $\ell$th partial wave in the $\pm$ channel. For the purposes of this analysis, we will only consider the ERPs up to order $k^{4}$. One can extend this analysis to include ERPs beyond this order by considering higher orders in the EFT.

\subsection{Power Counting For $s$-Waves and $p$-Waves}
Power counting is at the heart of the EFT framework. It allows us to systematically organize terms in the EFT expansion according to their relative importance. We define the dimensionless expansion parameter $Q \equiv \frac{k}{\Lambda_{B}}$. Here, $k$ is the momentum of the scattering process, and $\Lambda_{B}$ is the breakdown scale of the EFT. 

In this analysis, we will follow Poudel \& Phillips~\cite{Poudel_2022} and take $- \frac{2 k_{c} k^{2 \ell} H(\eta)}{e^{-\pi \eta} |\Gamma(1 + i\eta)|^{2}}$ (henceforth called the Coulomb characteristic component) to be the standard size of the inverse scattering amplitude. We will assign this piece to be of order $k_{c} k^{2 \ell}$. Our choice is motivated by the fact that the size of the Coulomb characteristic component is typically larger (as we will soon show) than the size of the other terms in the effective range function. 

A consequence of this choice is that no parameters enter the scattering amplitude at leading order (LO). Equivalently, at LO we have $a_{\ell}^{\pm} \rightarrow \infty$ and $r_{\ell}^{\pm}, P_{\ell}^{\pm} \rightarrow 0$. To determine the power counting of the ERPs, we then take the ratio of each term in the effective range function to the Coulomb characteristic component. This ratio provides us with a measure of the relative impact that each term has on the scattering amplitude, allowing us to develop a systematic power counting scheme. To illustrate this, we adopt the parameter values from the full $f$-wave analysis (discussed in Sec.~\ref{sec:implementation_with_f_waves} and listed here in Tab.~\ref{tab:f_wave_params}) and plot the ratios of the different terms in the effective range function to the Coulomb characteristic component.

\begin{table}[h]
    \centering
    \caption{\label{tab:f_wave_params} Maximum a posteriori (MAP) values of the ERPs from the full $f$-wave analysis. Parameters are directly sampled unless labeled with $^{*}$; parameters with $^{*}$ are derived. Details of this analysis are discussed in Sec.~\ref{sec:implementation_with_f_waves}.}
    \begin{tabular}{c|c || c|c}
        \hline
        Parameter & MAP Value & Parameter & MAP Value \\
        \hline
        $\frac{1}{a_{0}}$ (fm$^{-1}$) & 0.018 & $\frac{1}{a_{1}^{-}}$ (fm$^{-3}$) & 0.002$^{*}$ \\
        $r_{0}$ (fm) & 0.868 & $r_{1}^{-}$ (fm$^{-1}$) & -0.001$^{*}$ \\
        $\frac{1}{a_{1}^{+}}$ (fm$^{-3}$) & 0.006$^{*}$ & $P_{1}^{-}$ (fm) & 2.013 \\
        $r_{1}^{+}$ (fm$^{-1}$) & -0.091$^{*}$ & $\frac{1}{a_{3}^{+}}$ (fm$^{-7}$) & 0.015$^{*}$ \\
        $P_{1}^{+}$ (fm$^{-3}$) & 1.441 & $r_{3}^{+}$ (fm$^{-5}$) & 0.674 \\
        & & $P_{3}^{+}$ (fm$^{-3}$) & -4.268 \\
        \hline
    \end{tabular}
\end{table}

\begin{figure*}[ht]
	\centering
	\includegraphics[width = 0.95\textwidth]{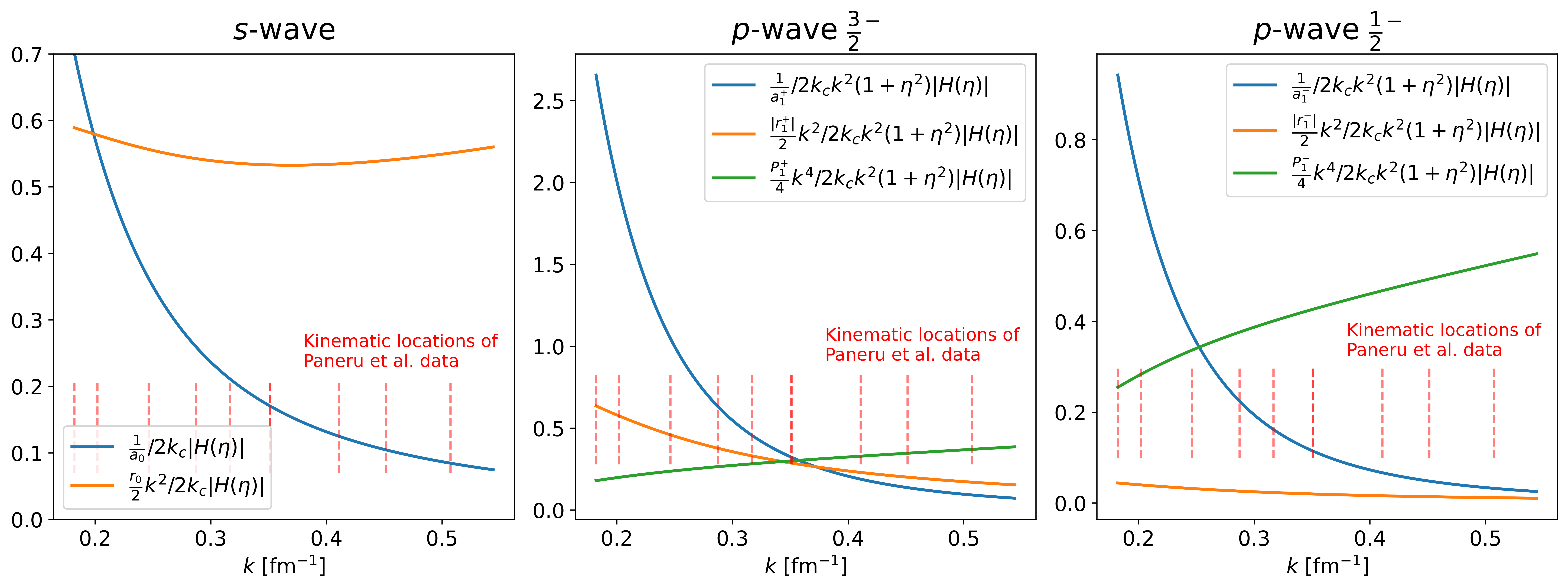}
	\caption{\label{fig:power_counting}Comparison of sizes of the different terms in the effective range function, relative to the Coulomb characteristic component of the CM-ERE. In each panel, the blue curve corresponds to the scattering length relative to the Coulomb characteristic component, the orange curve corresponds to the effective range relative to the Coulomb characteristic component, and the green curve corresponds to the shape parameter relative to the Coulomb characteristic component.}
\end{figure*}

Figure~\ref{fig:power_counting} shows the relative size of the different terms in the effective range function for the $s$-wave and $p$-wave channels across the kinematic region of interest where we have data ($0.18 \text{fm}^{-1} \le k \le 0.51 \text{fm}^{-1}$). To assign the power counting, we must consider both the size of these ratios, and also how the ratio changes across the kinematic region. We are particularly interested in the scaling at the kinematic locations where we compare to the data, labeled with the vertical dashed lines. These kinematic locations are where we will make inferences about the expansion coefficients of the EFT. 

For the $s$-wave, the scattering length $\frac{1}{a_{0}}$ is much smaller than the Coulomb characteristic component, though at low enough energies it becomes comparable. This parameter is unnaturally small when expressed in terms of the breakdown scale $\Lambda_{B}$, so we follow Poudel \& Phillips~\cite{Poudel_2022} and take its scaling to be $Q^{3} \Lambda_{B}$. Therefore we assign the $s$-wave scattering length $\frac{1}{a_{0}}$ to next-to-next-to leading order (N2LO). The effective range $r_{0}$ is approximately natural, and scales as $\frac{1}{\Lambda_{B}}$ so we assign it next-to-leading order (NLO) in the power counting. 

The $p$-wave scattering lengths are also unnaturally small and are assigned to N2LO. The effective ranges $r_{1}^{\pm}$ are surprisingly small and decrease in impact as the energy increases. We assign them to N2LO. The shape parameters $P_{1}^{\pm}$ increase in impact as the energy increases, and we assign them to NLO. This power counting scheme is similar to that used in Poudel \& Phillips~\cite{Poudel_2022}. Table~\ref{tab:heirarchy_within} summarizes the hierarchy of the $s$-wave and $p$-wave parameters in the EFT expansion.

\begin{figure*}[ht]
    \centering
    \includegraphics[width = 0.97\textwidth]{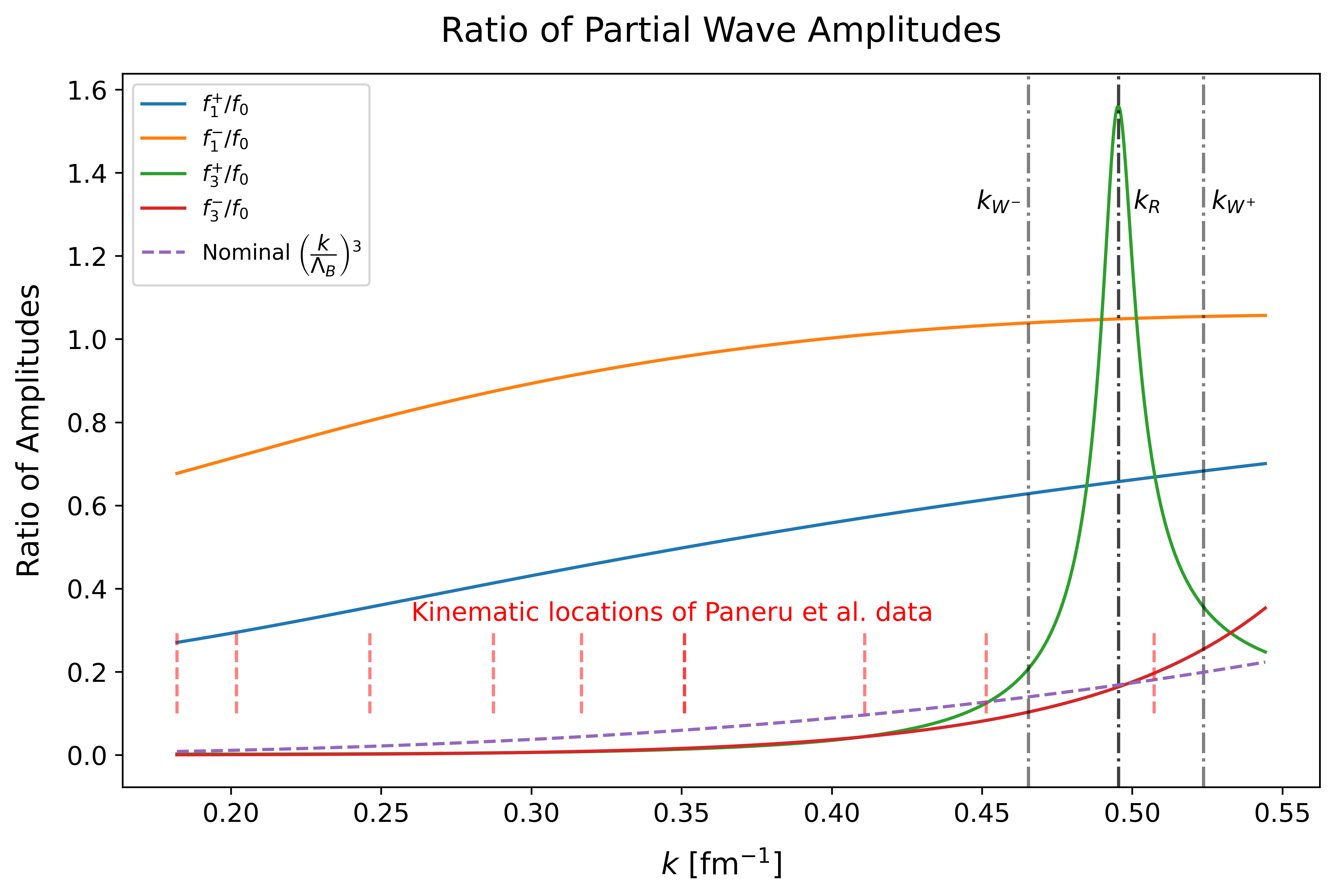}
    \caption{\label{fig:inv_amp_ratios}Comparison of the ratios of the amplitudes for the $^{3}$He-$\alpha$ system. The blue curve corresponds to the ratio of $\frac{3}{2}^{-}$ relative to $s$-wave, the orange corresponds to $\frac{1}{2}^{-}$ relative to $s$-wave, the green corresponds to $\frac{7}{2}^{-}$ relative to $s$-wave, and the red corresponds to $\frac{5}{2}^{-}$ relative to $s$-wave. The dashed purple line indicates the nominal $Q^{3}$ curve (with $\Lambda_B=0.90~{\rm fm}^{-1}$) indicating the size of N3LO effects. The short vertical dashed lines indicates the momentum at which we have cross section measurements in the Paneru dataset~\cite{Paneru, Paneru:thesis}. The tall vertical dashed-dotted lines indicate the resonance momentum $k_{R}$ and the window of the $\frac{7}{2}^{-}$ resonance. }
\end{figure*}

\subsection{Power Counting For $f$-Waves}
In order to more accurately describe the scattering data, we must incorporate the $f$-waves into our analysis. We include the $f$-waves by adding the $f_3^{\pm}$ amplitudes, parameterized through $\delta_{3}^{\pm}$ according to Eq.~(\ref{eq:f_ell_pm}). The $\frac{7}{2}^{-}$ resonance is a prominent feature, and it occurs in the $\delta_{3}^{+}$ channel at a momentum $k_{R} = \sqrt{2 \mu E_{R}} = 0.495 \text{ fm}^{-1}$, where $E_{R}$ is the real part of the resonance energy. Within the vicinity of the resonance, we must be able to reproduce the resonance position. This requires us to include at least both the $f$-wave scattering length $\frac{1}{a_{3}^{+}}$ and the $f$-wave effective range $r_{3}^{+}$. These two parameters allow the flexibility to place the real part of the resonance pole at $k_{R}$, therefore we will assign them to LO.

In the low energy region, far away from the resonance, the $f$-wave scattering amplitude is vastly suppressed when compared to the $s$- and $p$-wave scattering amplitudes. However, as we approach the resonance the $f$-wave scattering amplitude is enhanced and the power counting must be adjusted accordingly. Because of this, we will define a ``window'' around the resonance position in terms of the resonance width $\Gamma_{3}^{+}$. This window is defined as $k \in [k_{W^{-}}, k_{W^{+}}]$ where $k_{W^{\pm}} = \sqrt{2 \mu (E_{R} \pm 2 \Gamma_{3}^{+})}$, with resonance energy $E_{R}$. 

The green and red curves in Fig.~\ref{fig:inv_amp_ratios} show the ratios of the different $f$-wave inverse amplitudes relative to the $s$-wave inverse amplitude. In the low energy region ($k \ll k_{W^{-}}$), both $f$-wave channels are suppressed and considered to be higher order than N2LO. However, we include their effects in the EFT calculation there to ensure continuity of the scattering amplitude across the entire kinetic range studied here. We will eventually use part of this low energy region to estimate the size of the EFT expansion coefficients, however we will not include the $f$-waves in the inference of these coefficients.

\begin{figure}[h]
    \centering
    \includegraphics[width = 0.45\textwidth]{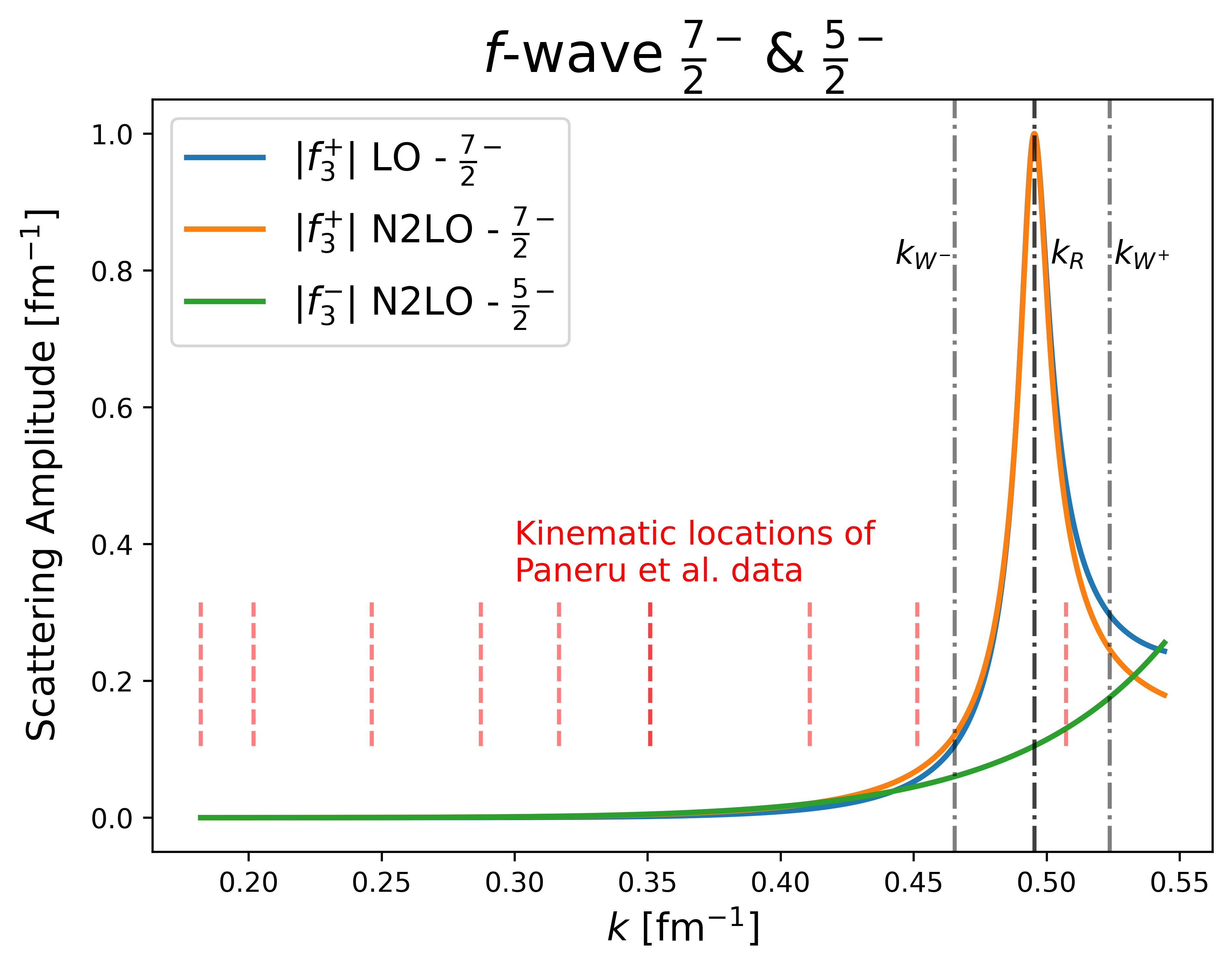}
    \caption{\label{fig:f_wave_power_counting} Comparison of the size of the $f$-wave scattering amplitudes at different orders of the EFT expansion. The vertical dashed lines indicate the $\frac{7}{2}^{-}$ resonance momentum $k_{R}$ and the window of the resonance.}
\end{figure}

The $f$-wave shape parameter $P_{3}^{+}$ is then a small correction to the scattering amplitude and is treated at N2LO. Fig.~\ref{fig:f_wave_power_counting} shows the different amplitudes of the $f$-wave channels. The N2LO amplitude includes $P_{3}^{+}$ and only has a small shift from the LO amplitude. 

The $\frac{5}{2}^{-}$ parameters are considered higher than N2LO, and are included to ensure continuity. The hierarchy including the $f$-waves within the resonance region is shown in Tab.~\ref{tab:heirarchy_within}.

\begin{table}[h]
	\centering
	\caption{\label{tab:heirarchy_within}Power counting hierarchy of the $s$-wave, $p$-wave, and $f$-wave parameters in the EFT expansion within the resonance window. The parameters enter the scattering amplitude at the order indicated in the table.}
	\begin{tabular}{c c c c}
		\hline
		Order & $s$-wave & $p$-wave & $f$-wave \\
		\hline
		LO & --- & --- & $\frac{1}{a_{3}^{+}}, r_{3}^{+}$ \\
		NLO & $r_{0}$ & $\frac{1}{a_{1}^{\pm}}, P_{1}^{\pm}$ & --- \\
		N2LO & $\frac{1}{a_{0}}$ & $r_{1}^{\pm}$ & $P_{3}^{+}$ \\
		\hline
	\end{tabular}
\end{table}

To quantify the theory uncertainty, we must consider the scaling of different components in the $f$-waves relative to the first omitted term in the effective range function. Our effective range function parameterization only goes up to $\mathcal{O}(k^{4})$, so we need to compare to the next term: $\frac{Q_{3}^{+}}{6}k^{6}$. 

We have two possible expansions to consider in the $\frac{7}{2}^{-}$ $f$-wave channel: one that works best at low energies, and then one that works best near the resonance. The expansion that works best at low energies is given by
\begin{align}
    \label{eq:f_wave_power_counting_low}
    k^{7} \big(\cot & \delta_{3}^{+} - i \big) = \frac{1}{C_{3}^{2}} \Bigg[ 36 \bigg( -\frac{1}{a_{3}^{+}} + \frac{r_{3}^{+}}{2} k^{2} + \frac{P_{3}^{+}}{4} k^{4} + \frac{Q_{3}^{+}}{6} k^{6} \bigg) \nonumber \\ 
    - & 2 k_{c} k^{6} (1 + \eta^{2}) (4 + \eta^{2}) (9 + \eta^{2}) H(\eta) \Bigg].
\end{align}

When expressed in terms of the resonance momentum $k_{R}$, we can write an analogous expression for the right-hand-side of Eq.~(\ref{eq:f_wave_power_counting_low}). This alternate expansion is better suited for energies near the resonance
\begin{align}
    \label{eq:f_wave_power_counting_near}
    & \frac{r_{3}^{+}}{2 C_{3}^{2}} \Bigg[ 36 \bigg( (k^{2} - k_{R}^{2}) + \frac{P_{3}^{+}}{2 r_{3}^{+}} (k^{4} - k_{R}^{4}) + \frac{Q_{3}^{+}}{3 r_{3}^{+}} (k^{6} - k_{R}^{6}) \bigg) \nonumber \\ 
    & - \frac{4 k_{c}}{r_{3}^{+}} \bigg( k^{6} V_{3}(\eta) H(\eta) - k_{R}^{6} V_{3}(\eta_{R}) Re \left\{ H(\eta_{R}) \right\} \bigg) \Bigg].
\end{align}
Details of how we obtain this alternate expansion are discussed later in Sec.~\ref{sec:scattering_model}. In these expressions we define $C_{3}^{2} = \frac{e^{2 \pi \eta} - 1}{2 \pi \eta} V_{3}(\eta)$ and $V_{3}(\eta) = (1 + \eta^{2}) (4 + \eta^{2}) (9 + \eta^{2})$. 

We assume that the higher order ERPs $r_{3}^{+}$, $P_{3}^{+}$, and $Q_{3}^{+}$ scale according to naive dimensional analysis: $r_{3}^{+} \sim \Lambda_{B}^{5}$, $P_{3}^{+} \sim \Lambda_{B}^{3}$, and $Q_{3}^{+} \sim \Lambda_{B}$. Additionally, we fine-tune the $f$-wave scattering length to reproduce the resonance position: $\frac{1}{a_{3}^{+}} \sim \Lambda_{B}^{5} k_{R}^{2}$. Using these scalings, we can then determine the relative error that accounts for the truncation of the effective range function in the two expansions:
\begin{align}
    \label{eq:f_wave_relative_error_below}
    \frac{Q_{3}^{+}}{6} k^{6} \Bigg/ \frac{1}{a_{3}^{+}} \sim \frac{\Lambda_{B} k^{6}}{6 \Lambda_{B}^{5} k_{R}^{2}} \nonumber \\ 
    = \frac{k^{6}}{6 \Lambda_{B}^{4} k_{R}^{2}} \hspace{0.75cm} \mbox{for $k < k_{W^-}$}
\end{align}
\begin{align}
    \label{eq:f_wave_relative_error_within}
    & \frac{Q_{3}^{+}}{3 r_{3}^{+}} (k^{6} - k_{R}^{6}) \Bigg/ (k^{2} - k_{R}^{2}) \sim \frac{\Lambda_{B}}{3 \Lambda_{B}^{5}} (k^{4} + k^{2} k_{R}^{2} + k_{R}^{4}) \nonumber \\
    &= \frac{1}{3 \Lambda_{B}^{4}} (k^{4} + k^{2} k_{R}^{2} + k_{R}^{4}) \hspace{0.2cm} \text{for } k_{W^{-}} \le k \le k_{W^{+}}.
\end{align}
These are the relative errors in the low energy region, and within the resonance region respectively. An important note is that the estimated error in Eq.~(\ref{eq:f_wave_relative_error_below}) is small when compared to the $s$ and $p$-wave relative errors. Ultimately these quantities will be used to construct the theory covariance matrix in Sec.~\ref{sec:uncertainty_model}.

We also note that the $P_{3}^{+}$ relative order follows a similar pattern to that of $Q_{3}^{+}$. The $P_{3}^{+}$ thus has a relative order of $\frac{k^{2}}{\Lambda_{B}^{2}}$ in the region near the resonance. This is consistent with our earlier assignment of $P_{3}^{+}$ to N2LO.

%% file: data.tex
 \section{Data}
\label{sec:data}
\begin{figure*}[t]
    \centering
    \includegraphics[width = 0.975\textwidth]{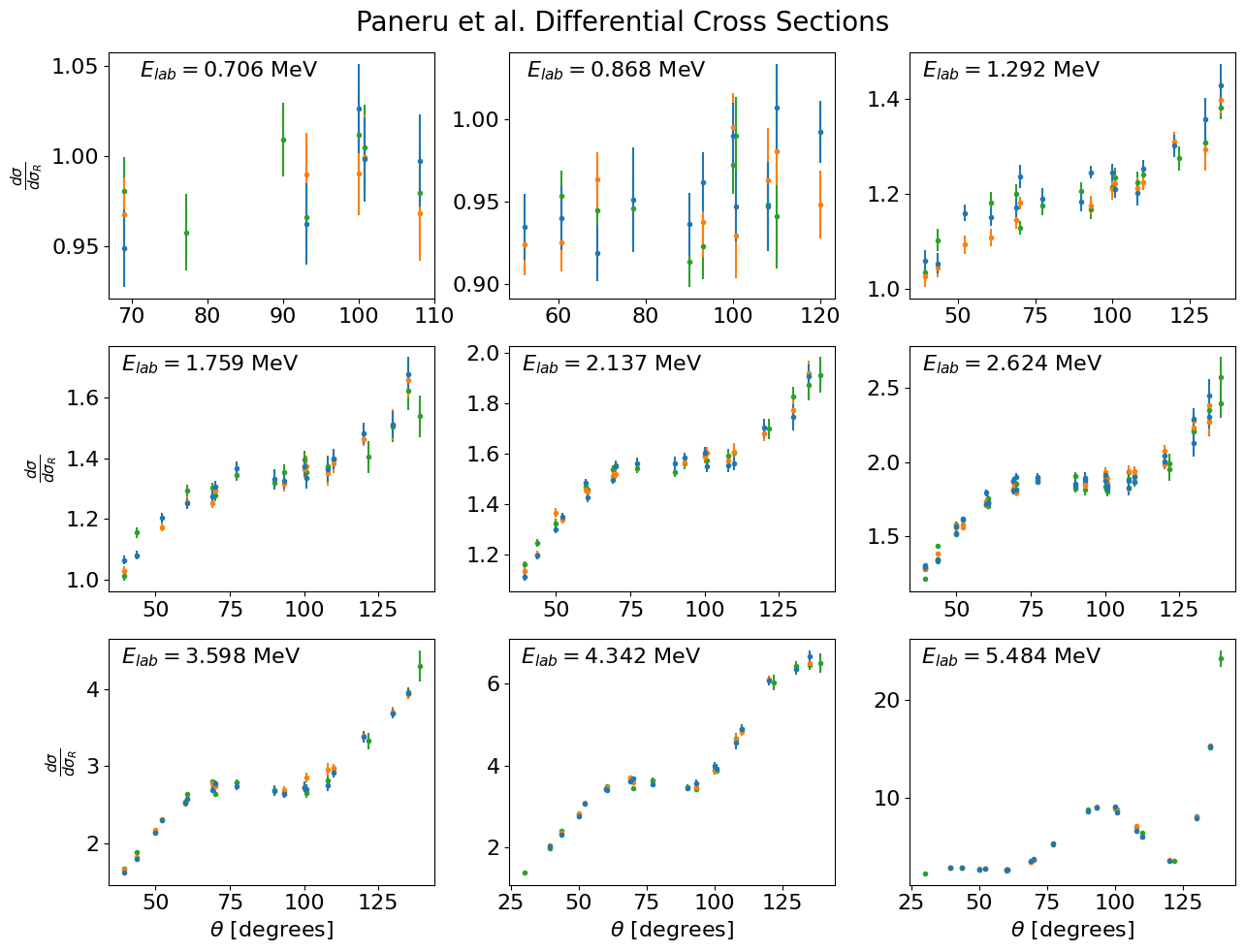}
    \caption{\label{fig:som_data}Differential cross section measurements of $^{3}$He-$^{4}$He elastic scattering from Paneru \textit{et al.}~\cite{Paneru}. Each panel corresponds to a different beam energy, indicated in the top left corner of each panel. The different colors correspond to the different interaction regions of the SONIK cell: blue is region 1, orange is region 2, and green is region 3. The panel corresponding to $E_{\text{lab}} = 2.6$ MeV contains data from two different runs, both taken at the same beam energy.}
\end{figure*}

\begin{table*}[ht]
    \caption{\label{tab:norm_priors}Total number of data points in each energy bin, and the standard deviations for each of the Gaussian priors on the normalization coefficients.}
    \centering
    \begin{tabular}{ccccccccccc}
        \toprule
        $E_{lab}$ (MeV) & 0.706 & 0.868 & 1.292 & 1.759 & 2.137 & 2.624 & 2.624 & 3.598 & 4.342 & 5.484 \\
        \toprule
        $N_{data}^{E}$ & 17 & 29 & 45 & 46 & 52 & 52 & 52 & 52 & 53 & 53 \\
        $\sigma_{\xi_{i}}$ & 0.064 & 0.076 & 0.098 & 0.057 & 0.045 & 0.062 & 0.041 & 0.077 & 0.063 & 0.089 \\
        \toprule
    \end{tabular}
\end{table*}

The dataset we use in this analysis is of the low-energy elastic scattering of $^{3}$He and $^{4}$He, from the experiment performed at TRIUMF by Paneru \textit{et al.}~\cite{Paneru}. This dataset will be referred to as the ``Paneru dataset''. The data consists of differential cross section measurements with $^{3}$He beam energies ranging from 0.7 to 5.5 MeV impinging on a $^{4}$He gas target chamber in the SONIK cell. The array of detectors span an angular range from $30^{\circ} < \theta_{cm} < 139^{\circ}$. The SONIK cell has 3 ``interaction'' regions where the scattering occurs. Due to the different interaction locations, measurements at different interaction regions have slightly different energies. This difference is taken into account in the analysis, but for plotting purposes we will plot based on incoming beam energies. The scattering experiment was performed at 10 different beam energies; the measurement at $E_{lab} = 2.6$ MeV was performed twice. 

The measured cross sections and experimental uncertainties are depicted in Fig.~\ref{fig:som_data}. Full details of the experimental setup and data analysis can be found in the original publication~\cite{Paneru}. The data and uncertainties are reported in multiple tables at the end of Paneru's dissertation~\cite{Paneru:thesis}. 

Each set of measurements at a given beam energy has an overall normalization factor associated with it. These normalization factors account for the common mode error in a particular energy bin. These parameters will be denoted as $\bm{\xi} = \{ \xi_{1}, \xi_{2}, \ldots, \xi_{N_{E}} \}$ where $N_{E}=10$ is the number of energy bins. 

We define $N_{\text{data}}$ as the total number of data points included in the analysis. Here we consider different subsets of the data: the first consideration is data with $E_{lab} \le 2.6$ MeV, next we have data with $E_{lab} \le 3.6$ MeV, and then lastly all data reported from~\cite{Paneru}. 

In addition to the Paneru dataset, we also consider phase shift extractions for the $\frac{5}{2}^{-}$ channel from Boykin \textit{et al.}~\cite{Boykin:1972}. This phase shift extraction uses cross section data obtained from Barnard \textit{et al.}~\cite{Barnard:1964} and Spiger \& Tombrello~\cite{Spiger:1967}. The Boykin phase shift extractions and knowledge of the $^{7}$Be resonance state at $E = 6.73$ MeV~\cite{Tilley2002} allow us to fit the $\frac{5}{2}^{-}$ channel ERPs $a_{3}^{-}$ and $r_{3}^{-}$.

%% file: scattering_model.tex
\section{Scattering Model}
\label{sec:scattering_model}
\subsection{Differential Cross Section}
The differential cross section is given by the sum of the squared scattering amplitudes
\begin{equation}
    \label{eq:diff_cross}
    \frac{d\sigma}{d\Omega} = |f_{c}|^{2} + |f_{i}|^{2}.
\end{equation} 
It is useful to decompose these amplitudes into partial waves~\cite{Spiger_1967,Critchfield:1949}. The partial-wave decomposition is given by
\begin{align}
    \label{eq:fc}
    f_{c} = - \frac{\eta}{2 k} \csc^{2}(\theta / 2) & \exp\left( i \eta \log(\csc^{2}(\theta / 2)) \right) \nonumber \\ + \frac{1}{k} \sum_{\ell = 0}^{\infty} & \exp(2 i \alpha_{\ell}) P_{\ell}(\cos \theta) \nonumber \\ \times & \left[ \frac{\ell + 1}{\cot \delta_{\ell}^{+} - i} + \frac{\ell}{\cot \delta_{\ell}^{-} - i} \right],
\end{align}
\begin{align}
    \label{eq:fi}
    f_{i} = \frac{1}{k} \sum_{\ell = 0}^{\infty} \exp(2 i \alpha_{\ell}) & \sin \theta \frac{d P_{\ell}(\cos \theta)}{d \cos \theta} \nonumber \\ \times & \left[ \frac{1}{\cot \delta_{\ell}^{-} - i} - \frac{1}{\cot \delta_{\ell}^{+} - i} \right].
\end{align}
The first term in Eq.~(\ref{eq:fc}) is the Rutherford scattering amplitude ($f_{R}$). Meanwhile, $\alpha_{\ell}$ is the difference in Coulomb phase shifts, $\alpha_{\ell} = \sigma_{\ell} - \sigma_{0}$, $\delta_{\ell}^{\pm}$ are the $\ell$th phase shifts of the $\pm$ channel (corresponding to total angular momentum state $j = \ell \pm \frac{1}{2}$), $P_{\ell}$ are the Legendre polynomials, $\theta$ is the scattering angle in the center-of-mass frame, $k$ is the wave number, and $\eta$ is the Sommerfeld parameter. 

We analyze bound and resonance states in our scattering model by analytically continuing the scattering momenta to complex values. The bound states appear as poles with momentum $k_{B} = i \gamma_{B}$ (with corresponding complex $\eta_{B} = -i \frac{k_{c}}{\gamma_{B}}$), where $\gamma_{B}$ is the real binding momenta. Resonance states appear as poles in the scattering amplitude. The real part of the momentum associated with the pole can be denoted $k_{R}$ (with corresponding $\eta_{R} = \frac{k_{c}}{k_{R}}$). The imaginary part of the resonance pole is associated with the width of the resonance. We can include these bound and resonance states by enforcing the condition that at the appropriate resonance or binding momentum in the relevant channel, the real part of the quantity in square brackets in Eq.~(\ref{eq:invamp}) is zero. This condition is applied most easily through the channel's scattering length.

\subsection{S \& P Waves}
For the $\frac{3}{2}^{-}$ and $\frac{1}{2}^{-}$ $p$-wave states (corresponding to the $+$ and $-$ superscript of the phase shifts with $\ell = 1$ respectively), this yields the following conditions on the $p$-wave scattering lengths
\begin{align}
    \label{eq:p_wave_condition}
    \frac{1}{a_{1}^{\pm}} = - \frac{r_{1}^{\pm}}{2} & (\gamma_{B}^{\pm})^{2} + \frac{P_{1}^{\pm}}{4} (\gamma_{B}^{\pm})^{4} \nonumber \\
    & + 2 k_{c} (\gamma_{B}^{\pm})^{2} (1 + \eta_{B}^{2}) \Re \left[ H(-i \eta_{B}) \right].
\end{align}
Expressing the $p$-wave scattering lengths in this way benefits us in two ways: first, it ensures that the bound state pole is in the correct position in the scattering amplitude, and second, it allows us to infer $a_{1}^{\pm}$ through $r_{1}^{\pm}$ and $P_{1}^{\pm}$ thus reducing the sampling dimension.

We can also express the $p$-wave effective range $r_{1}^{\pm}$ in terms of the square of the asymptotic normalization coefficients (ANCs) $(C_{1}^{\pm})^{2}$. The effective range is related to the ANC~\cite{Poudel_2022} via
\begin{align}
    \label{eq:ANC_relation}
    r_{1}^{\pm} = \frac{- 2 (\gamma_{B}^{\pm})^{2} \Gamma(2 + \eta_{B}^{\pm})^{2}}{(C_{1}^{\pm})^{2}} + P_{1}^{\pm} & (\gamma_{B}^{\pm})^{2} + 4 k_{c} H(-i \eta_{B}^{\pm}) \nonumber \\
    + 2 i k_{c} \eta_{B}^{\pm} (1 - (\eta_{B}^{\pm})^{2}) & \frac{d H(\eta)}{d \eta} \Bigg \vert_{\eta = -i \eta_{B}^{\pm}}.
\end{align}
Since the ANC is directly related to the solution of the radial Schrödinger equation for the corresponding bound state, we have better knowledge of the ANCs than of the $p$-wave effective ranges. Hence, we can use our prior knowledge of the ANCs to provide a more accurate estimate of the $p$-wave ERPs. This prior information is discussed in more detail in Sec.~\ref{sec:simultaneous_sampling}.

\subsection{F Waves and Resonances}
A resonance has two important properties: the real part of the resonance pole's momentum $k_{R}$ and the width of the resonance $\Gamma_{R}$. 

For momentum $k=k_{R}$ (with associated $\eta_{R} = \frac{k_{c}}{k_{R}}$), the real part of the quantity in square brackets in Eq.~(\ref{eq:invamp}) is zero. This condition allows us to tune the $f$-wave scattering length $\frac{1}{a_{3}^{+}}$ to reproduce the real part of the resonance pole position, giving us the following constraint
\begin{align}
    \label{eq:resonance_condition}
    \frac{1}{a_{3}^{+}} = \frac{r_{3}^{+}}{2} k_{R}^{2} + \frac{P_{3}^{+}}{4} k_{R}^{4} - \Bigg( \frac{2 k_{c} k_{R}^{6}}{36}  (1 + \eta_{R}^{2}) \nonumber \\ \times (4 + \eta_{R}^{2}) (9 + \eta_{R}^{2}) \Re \left[ H(\eta_{R}) \right] \Bigg).
\end{align}

The resonance width is related to the derivative of the scattering amplitude evaluated at the resonance momentum $k_{R}$ via
\begin{equation}
    \label{eq:width}
    \Gamma_{3}^{+} = \frac{- k^{7} C_{3}^{2}}{\mu \tau_{3}^{+}},
\end{equation}
with $\tau_{3}^{+}$ defined as
\begin{align}
    \label{eq:tau_3}
    \tau_{3}^{+} = 36 \left( \frac{r_{3}^{+}}{2} + \frac{P_{3}^{+}}{3} k_{R}^{2} \right) - \frac{d}{d k^{2}} \Bigg[ 2 k_{c} k^{6} (1 + \eta^{2}) \nonumber \\ 
    \times (4 + \eta^{2}) (9 + \eta^{2}) \Re \left\{ H(\eta) \right\} \Bigg] \Bigg|_{k = k_{R}}.
\end{align}
The conditions in Eqs.~(\ref{eq:width}-\ref{eq:tau_3}) are derived and further discussed in Appendix~\ref{app:width}. (Note that the parameter $C_{3}^{2}$ that appears in Eq.~(\ref{eq:width}) is the Coulomb penetration factor, and not the ANC.) These equations allow us to relate the $f$-wave ERPs to the resonance position and width. They are valid assuming we take the $f$-wave CM-ERE up to $\mathcal{O}(k^{4})$.

%% file: uncertainty_model.tex
\section{Uncertainty Model}
\label{sec:uncertainty_model}
The uncertainty models in previous EFT analyses of the $^{3}$He-$\alpha$ system have been constructed at the cross section level~\cite{Poudel_2022, Burnelis:2024}. While this approach is useful and provides predictive power, it lacks the flexibility to easily account for changing power counting schemes across kinematic regions. While Melendez \textit{et al.}~\cite{Melendez:2021} did account for a changing power counting scheme in the context of proton Compton scattering, their work also used the theory covariance matrix at the cross section level.

In this work, we instead choose to implement the uncertainty at the level of the contributions of the partial wave amplitudes to the scattering amplitude. By using this decomposition we are able to accommodate the changing power counting scheme across the kinematic regions, which is particularly important in this system due to the presence of the $\frac{7}{2}^{-}$ resonance.

We implement this uncertainty model by further decomposing the scattering amplitudes appearing in Eq.~(\ref{eq:diff_cross}) into relevant partial waves. The non-spin-flip and spin-flip amplitudes are decomposed as follows:
\begin{align}
	\label{eq:f_decomp}
	f_{c} & = f_{R} + f_{c}^{(0)} + \delta f_{c}^{(0)} + f_{c}^{(1)} + \delta f_{c}^{(1)} + f_{c}^{(3)} + \delta f_{c}^{(3)}, \nonumber \\
	f_{i} & = f_{i}^{(1)} + \delta f_{i}^{(1)} + f_{i}^{(3)} + \delta f_{i}^{(3)}.
\end{align}
The terms $\delta f_{c}^{(\ell)}$ and $\delta f_{i}^{(\ell)}$ represent the uncertainties in the non-spin-flip and spin-flip amplitudes for the $\ell$th partial wave.
The term $f_{R}$ is the Rutherford amplitude which has no uncertainty associated with it. The terms $f_{c}^{(\ell)}$ and $f_{i}^{(\ell)}$ are the non-spin-flip and spin-flip amplitudes for the $\ell$th partial wave, defined, respectively, as the terms of the sums in Eqs.~(\ref{eq:fc}) \& (\ref{eq:fi}). Each $f$ appearing in Eq.~(\ref{eq:f_decomp}) contains both channels within the same partial wave, and the corresponding Legendre polynomials. All the $f$'s appearing in Eq.~(\ref{eq:f_decomp}) are to be evaluated at the highest order of the EFT expansion, N2LO $(n = 2)$. The amplitude at a given order $n$ is then defined as
\begin{equation}
	\label{eq:f_order}
	f_{x, n}^{(\ell)} = f_{x, 0}^{(\ell)} \sum_{j = 0}^{n} c_{x, j}^{(\ell)} \left( \frac{k}{\Lambda_{B}} \right)^{j},
\end{equation}
where $x = c, i$, $f_{x, 0}^{(\ell)}$ is the leading order amplitude, and $c_{x, j}^{(\ell)}$ are the coefficients of the EFT expansion. We take the leading order coefficient $c_{x, 0}^{(\ell)} = 1$ for all $x$ and $\ell$.

\subsection{Covariance Matrix}
As discussed previously in Sec.~\ref{sec:eft}, we have two different power counting schemes for the $f$-waves. To account for this, we define a piecewise covariance matrix that captures the different uncertainties in each region. Based on the power counting from Sec.~\ref{sec:eft}, we will assume the following distributions for the different uncertainties in the scattering amplitudes:
\begin{flalign}
	\label{eq:uncertainty_distribution}
	& \delta f_{c}^{(0)} \sim \mathcal{N} \left(0, {\left[\bar{c} Q^{3} f_{c}^{(0)} \right]^{2}} \right), \nonumber \\ 
	& \delta f_{x}^{(1)} \sim \mathcal{N} \left( 0, {\left[\bar{c} Q^{3} f_{x}^{(1)} \right]^{2}} \right), \nonumber \\
	& \delta f_{x, below}^{(3)} \sim \mathcal{N} \left( 0, {\left[\bar{c} {\left(\frac{k^{6}}{6 \Lambda_{B}^{4} k_{R}^{2}}\right)} f_{x}^{(3)} \right]^{2}} \right), \nonumber \\
	& \delta f_{x, in}^{(3)} \sim \mathcal{N} \left( 0, {\left[\bar{d} {\left( \frac{(k^4 + k^2 k_{R}^{2} + k_{R}^{4})}{3 \Lambda_{B}^{4}}\right)} f_{x}^{(3)} \right]^{2}} \right), 
\end{flalign}
where the subscript $x$ can be either $c$ or $i$, while $\bar{c}$ and $\bar{d}$ are the standard deviation of the expansion coefficients in their respective regions. Equation~(\ref{eq:uncertainty_distribution}) defines the standard deviations of each amplitude at momentum $k$, and helps us determine the covariance matrix.

Since the scattering amplitude is continuous across the resonance window boundary, we assume that the uncertainty is also continuous across the boundary. Matching at the boundary $k_{W-}$ leads to the following constraint on the expansion coefficient $\bar{d}$:
\begin{equation}
	\bar{d} = \frac{\bar{c} k_{W-}^{6}}{2 k_{R}^{2} (k_{W-}^{4} + k_{W-}^{2} k_{R}^{2} + k_{R}^{4})}.
\end{equation}

We can then express the full theory uncertainty in the differential cross section by propagating the uncertainties of the scattering amplitudes in Eq.~(\ref{eq:uncertainty_distribution}) through the differential cross section formula in Eq.~(\ref{eq:diff_cross}). Dropping all terms that are $\mathcal{O}\left[(\delta f_{x}^{(\ell)})^{2}\right]$ allows us to write the standard deviations in the differential cross section at the center of mass momuntum $k_{j}$ and scattering angle $\theta_{n}$ as:
\begin{widetext}
	\begin{eqnarray}
		\label{eq:delta_sigma}
		\Delta \sigma_{jn} = 
		\begin{cases}
			2 \bar{c} \Re \biggl\{ f_{c}^{*} \left[ Q^{3} (f_{c}^{(0)} + f_{c}^{(1)}) + \left(\frac{k^{6}}{6 \Lambda_{B}^{4} k_{R}^{2}}\right) f_{c}^{(3)}  \right] + f_{i}^{*} \left[ Q^{3} f_{i}^{(1)} + \left(\frac{k^{6}}{6 \Lambda_{B}^{4} k_{R}^{2}}\right) f_{i}^{(3)} \right] \biggr\}_{j,n}, & k_{j} < k_{W-}, \\
			2 \bar{c} \Re \biggl\{ f_{c}^{*} \left[ Q^{3} (f_{c}^{(0)} + f_{c}^{(1)}) + \frac{k_{W-}^{6}}{2 k_{R}^{2} (k_{W-}^{4} + k_{W-}^{2} k_{R}^{2} + k_{R}^{4})} \left( \frac{k^{4} + k^{2} k_{R}^{2} + k_{R}^{2}}{3 \Lambda_{B}^{4}} \right) f_{c}^{(3)} \right] \\ 
			\quad \quad \quad \quad \quad + f_{i}^{*} \left[ Q^{3} f_{i}^{(1)} + \frac{k_{W-}^{6}}{2 k_{R}^{2} (k_{W-}^{4} + k_{W-}^{2} k_{R}^{2} + k_{R}^{4})} \left( \frac{k^{4} + k^{2} k_{R}^{2} + k_{R}^{2}}{3 \Lambda_{B}^{4}} \right) f_{i}^{(3)} \right] \biggr\}_{j,n}, & k_{W-} \leq k_{j} \leq k_{W+}. \\
		\end{cases}
	\end{eqnarray}
\end{widetext}
We note that all angular dependence is implicitly contained within the scattering amplitudes $f_{c}^{(\ell)}$ and $f_{i}^{(\ell)}$. The uncertainties $\Delta \sigma_{jn}$ thus can be interpreted as a vector of uncertainties. We assume a fully correlated theory uncertainty so we then construct the theory covariance matrix by taking the outer product of the vector of uncertainties with itself:
\begin{equation}
	\label{eq:theory_covariance}
	\Sigma_{jn, km}^{th} = \Delta \sigma_{jn} \Delta \sigma_{km}.
\end{equation}

Conversely, we assume the experimental covariance matrix to be diagonal and defined as
\begin{equation}
	\Sigma_{jn, km}^{exp} = \sigma_{j, n} \sigma_{k, m} \delta_{jk} \delta_{nm},
\end{equation}
where $\sigma_{j, n}$ $(\sigma_{k, m})$ (not to be confused with the cross section) is the experimental point-to-point uncertainty in the $j$th $(k$th) energy and $n$th $(m$th) angle data point. It is worth noting here that we may easily accommodate more complex experimental covariance structures. To construct the total covariance used in the likelihood, we simply add the theory and experimental covariance matrices together:
\begin{equation}
	\label{eq:full_covariance}
	\Sigma_{jk} = \Sigma_{jk}^{th} + \Sigma_{jk}^{exp}.
\end{equation}

%% file: simultaneous_sampling.tex
\section{Simultaneously Sampling Model Parameters and Truncation Uncertainty Parameters}
\label{sec:simultaneous_sampling}
We utilize a Bayesian approach to simultaneously sample the model parameters and the truncation uncertainty parameters. To do this, we use Bayes' theorem to construct the desired joint posterior distribution
\begin{equation}
	\label{eq:joint_posterior}
	\hspace*{-0.945em} P(\boldsymbol{\theta}, \bar{c}^{2}, \Lambda_{B} | D, I) = \frac{P(D | \boldsymbol{\theta}, \bar{c}^{2}, \Lambda_{B}, I) P(\boldsymbol{\theta}, \bar{c}^{2}, \Lambda_{B} | I)}{P(D | I)}.
\end{equation}
Here we adopt the shorthand notation for the model parameters, $\boldsymbol{\theta} = \{ \boldsymbol{a}, \bm{\xi} \}$ where 
\begin{equation}
	\label{eq:a_params}
    \boldsymbol{a} = \Bigg\{A_{0}, r_{0}, (C_{1}^{+})^{2}, P_{1}^{+}, (C_{1}^{-})^{2}, P_{1}^{-} \Bigg\},
\end{equation}
are the physics model parameters (composed of ERPs and the ANCs), and $\bm{\xi}$ are the cross section normalization factors as discussed in Sec.~\ref{sec:data}. The parameter $A_{0} \equiv \frac{1}{a_{0}}$, is the inverse of the $s$-wave scattering length. We define $N_{\text{params}}$ as the number of physics model parameters. In the case without $f$-waves, we have $N_{\text{params}} = 6$, and when we include the $f$-waves, we augment the set of physics model parameters to include $r_{3}^{+}$ and $P_{3}^{+}$, giving us $N_{\text{params}} = 8$.

In many cases, the model evidence (sometimes called the marginal likelihood), $P(D | I)$ is not needed as it only sets an overall scaling factor for the posterior distribution. The evidence is independent of any parameters and does not affect the general shape of the distribution with respect to the parameters. We therefore rewrite:
\begin{align}
    \label{eq:joint_posterior_expanded}
    P(\boldsymbol{\theta}, \bar{c}^{2}, \Lambda_{B} | D, I) \propto P(D | \boldsymbol{\theta}, \bar{c}^{2}, \Lambda_{B}, I) \nonumber \\ 
	\times P(\bar{c}^{2}, \Lambda_{B} | \boldsymbol{\theta}, I) P(\boldsymbol{\theta} | I),
\end{align}
where we have further expanded the prior. Here, $P(D | \boldsymbol{\theta}, \bar{c}^{2}, \Lambda_{B}, I)$ is the likelihood, $P(\bar{c}^{2}, \Lambda_{B} | \boldsymbol{\theta}, I)$ is the ``joint intermediate-prior'' distribution of the truncation uncertainty parameters, and $P(\boldsymbol{\theta} | I)$ is the prior on the model parameters. 

The likelihood function is defined as
\begin{equation}
    \label{eq:likelihood}
    P(D | \boldsymbol{\theta}, \bar{c}^{2}, \Lambda_{B}, I) = \frac{1}{\sqrt{(2 \pi)^{N_{\text{data}}} \det \Sigma}} \exp \left( - \frac{\chi^{2}}{2} \right),
\end{equation}
with $\Sigma$ being the total covariance matrix as defined in Eq.~(\ref{eq:full_covariance}) and 
\begin{align}
    \chi^{2} = \sum_{j, k = 1}^{N_{E}} \sum_{n = 1}^{N_{E_{j}, \text{data}}} & \sum_{m=1}^{{E_{k}, \text{data}}} \left( y_{j, n} - {\xi}_{j} y(E_{j}, \theta_{n}, \boldsymbol{a}) \right) [\Sigma]_{jn, km} \nonumber \\ 
	\times &\left( y_{k, m} - {\xi}_{k} y(E_{k}, \theta_{m}, \boldsymbol{a}) \right).
\end{align}
Here, $N_{E_{j}, \text{data}}$ is the number of data at energy $E_{j}$, $y_{j, n}$ is the $n$th data point at energy $E_{j}$, $\xi_{j}$ is the normalization for the $j$th energy bin, and $y(E_{j}, \theta_{n}, \boldsymbol{a})$ is the model prediction at energy $E_{j}$, angle $\theta_{n}$, and physics model parameters $\boldsymbol{a}$. 

We split the prior on the model parameters into two parts: the physics model parameters and the normalization coefficients
\begin{equation}
	\label{eq:prior_expanded}
	P(\boldsymbol{\theta} | I) = P(\boldsymbol{a} | I) P(\bm{\xi} | I).
\end{equation}
The priors on $\boldsymbol{a}$ and $\bm{\xi}$ are both truncated Gaussian distributions~\cite{Poudel_2022, Burnelis:2024} defined as
\begin{eqnarray}
	\label{eq:a_prior}
	P(\boldsymbol{a} | I) = \prod_{i = 1}^{N_{\text{params}}} \mathcal{N}(\mu_{i}, \sigma_{i}^{2}) T(a_{i}, b_{i}), \nonumber \\
	P(\bm{\xi} | I) = \prod_{i = 1}^{N_{E}} \mathcal{N}(1, \sigma_{\xi_{i}}^{2}) T(0, 2).
\end{eqnarray}
Here we have made use of $T(a, b)$ as the truncated uniform distribution defined as
\begin{equation}
	T(a, b) = \begin{cases}
	1 & \text{if } [a, b] \\
	0 & \text{otherwise}.
	\end{cases}
\end{equation}
The values of the means, variances, and truncation limits for the physics model parameters are defined in Tab.~\ref{tab:param_priors1}. Analysis of $^{3}$He-$^{4}$He capture data by Zhang~\textit{et al.}~\cite{Zhang:2020} inform the priors for the $p$-wave ANCs, ${(C_{1}^{+})}^{2}$ and ${(C_{1}^{-})}^{2}$. Their analysis also found a large positive $s$-wave scattering length $\frac{1}{a_{0}}$, which is consistent with the prior chosen here. The $s$-wave effective range $r_{0}$ is positive, and the prior is chosen to be consistent with the analysis of Poudel \& Phillips~\cite{Poudel_2022}. The priors for the $p$-wave shape parameters $P_{1}^{\pm}$ and the $f$-wave parameters are chosen to be broad and consistent with naturalness. The standard deviations for the normalization coefficients are defined in Tab.~\ref{tab:norm_priors}, based on information from Ref.~\cite{Paneru}.

\begin{table}[h]
    \caption{\label{tab:param_priors1}Truncation bounds, means, and variances of the Gaussian priors for each of the physics model parameters. Equation~(\ref{eq:a_prior}) shows the full prior on the effective range parameters. Values are listed here in the appropriate units of $\text{fm}^{n}$ as they are provided in Tab.~\ref{tab:f_wave_params}.}
    \centering
    \begin{tabular}{lcccccccc}
        \toprule
        Parameter & $\frac{1}{a_{0}}$ & $r_{0}$ & ${(C_{1}^{+})}^{2}$ & $P_{1}^{+}$ & ${(C_{1}^{-})}^{2}$ & $P_{1}^{-}$ & $r_{3}^{+}$ & $P_{3}^{+}$ \\
        \toprule
        $a$ & -0.02 & -3.0 & 5.0 & -6.0 & 5.0 & -6.0 & -6.0 & -6.0 \\
        $b$ & 0.06 & 3.0 & 25.0 & 6.0 & 25.0 & 6.0 & 6.0 & 6.0 \\
        $\mu$ & 0.025 & 0.8 & 13.84 & 0.0 & 12.59 & 0.0 & 0.0 & 0.0 \\
        $\sigma$ & 0.015 & 0.4 & 1.63 & 1.6 & 1.85 & 1.6 & 1.0 & 1.0 \\
        \toprule
    \end{tabular}
\end{table}

Now we construct the joint intermediate-prior distribution of the truncation uncertainty parameters, $P(\bar{c}^{2}, \Lambda_{B} | \boldsymbol{\theta}, I)$. We follow the logic of Melendez \textit{et al.} and Wesolowski \textit{et al.}~\cite{Melendez:2019, Wesolowski:2021}. The first step is to expand the joint intermediate-prior distribution as
\begin{equation}
    \label{eq:intermediate_prior}
	P(\bar{c}^{2}, \Lambda_{B} | \boldsymbol{\theta}, I) = P(\bar{c}^{2} | \boldsymbol{\theta}, \Lambda_{B}, I) P(\Lambda_{B} | \boldsymbol{\theta}, I).
\end{equation}

We first focus on the first term, $P(\bar{c}^{2} | \boldsymbol{\theta}, \Lambda_{B}, I)$. This piece is the distribution of $\bar{c}^{2}$ based on the size of the expansion coefficients, $\boldsymbol{c} = \left\{ c_{n} \right\}$ determined by the different orders of the EFT calculation. We assume that the expansion coefficients are Gaussian distributed with variance $\bar{c}^{2}$, and that $\bar{c}^{2}$ is distributed according to a scaled inverse chi-squared distribution~\cite{Melendez:2019, Wesolowski:2021}
\begin{eqnarray}
	\label{eq:bar_c_prior}
	P(\boldsymbol{c} | \bar{c}^{2}, I) = \mathcal{N}(0, \bar{c}^{2}), \nonumber \\
	P(\bar{c}^{2} | I) = \chi^{-2}\left[ \nu_{0}, \tau_{0}^{2} \right].
\end{eqnarray}
The choice to use a scaled inverse chi-squared distribution is motivated by two factors: first, it behaves in a desirable way preventing $\bar{c}^{2}$ from becoming too small or too large, and second, it is conjugate to the Gaussian distribution of the expansion coefficients~\cite{Melendez:2019, Wesolowski:2021}. We choose $\nu_{0} = 1.5$ (initial degrees of freedom) and $\tau_{0}^{2} = 1.5$ (initial scale) as the hyperparameters that define the prior on $\bar{c}^{2}$. 

The set of coefficients $\boldsymbol{c}$ are uniquely determined by the physics model parameters $\boldsymbol{\theta}$ and the breakdown scale $\Lambda_{B}$. Therefore, we will do a change of variables via marginalization to express the intermediate-prior on $\bar{c}^{2}$ in terms of the coefficients $\boldsymbol{c}$ 
\begin{align}
	\label{eq:marginalization}
	P(\bar{c}^{2} | \boldsymbol{\theta}, \Lambda_{B}, I) = \int P(\bar{c}^{2} | \boldsymbol{c}, \boldsymbol{\theta}, & \Lambda_{B}, I) \nonumber \\
	\times & P(\boldsymbol{c} | \boldsymbol{\theta}, \Lambda_{B}, I) d\boldsymbol{c}.     
\end{align}
Since $\boldsymbol{c}$ is a deterministic function of $\boldsymbol{\theta}$ and $\Lambda_{B}$, we can treat $P(\boldsymbol{c} | \boldsymbol{\theta}, \Lambda_{B}, I)$ as a delta function and perform this integral analytically. The delta function is defined to select a discrete set of $\boldsymbol{c}$ values, which are evaluated at energies where we have data.

The choice of which kinematic points and which partial waves to include in the set of $\boldsymbol{c}$ values is important as it determines the convergence pattern of the EFT expansion. We consider the kinematic points that are between the low momentum $k = 0.34$ fm$^{-1}$ and the resonance region that starts at $k = 0.468$ fm$^{-1}$. These points are chosen because, at momenta lower than $0.34$ fm$^{-1}$, Coulomb interactions dominate the scattering amplitude. On the other hand, at momenta above $0.468$ fm$^{-1}$ we are in the resonance region. We also do not consider the $f$-wave coefficients because their contribution to the scattering amplitude is negligible below the resonance region (see Fig.~\ref{fig:inv_amp_ratios}). The $p$-wave spin-flip coefficients, $c_{i, n}^{(1)}$ are not included because $f_{i, n}^{(1)}$ has a pole at $90^{\circ}$ which causes instability when extracting coefficients. Hence, for each data point we include in the inference of $\bar{c}^{2}$, we obtain 4 coefficients: NLO \& N2LO for the $s$-wave amplitude and NLO \& N2LO for the $p$-wave spin-preserving amplitude. Our final set of coefficients is then 
\begin{equation}
	\label{eq:coefficients_set}
	\boldsymbol{c} = \left\{ c^{(0)}_{c, 1, j}, c^{(0)}_{c, 2, j}, c^{(1)}_{c, 1, j}, c^{(1)}_{c, 2, j} \right\}.
\end{equation}
Each coefficient follows the notation $c_{c, n, j}^{(\ell)}$ where the superscript $\ell$ indicates the partial wave, the subscript $c$ indicates that this is the non-spin-flip coefficient, the subscript $n$ indicates the order in the EFT expansion, and the subscript $j$ indexes the energy bin where we evaluate the coefficient. The index $j$ runs over the energy bins with data between $0.34$ fm$^{-1}$ and $0.468$ fm$^{-1}$, i.e. $j \in \{1, \ldots, N_{in} \}$ where $N_{in}$ is the total number of energy bins with data in this range. The total number of coefficients in $\boldsymbol{c}$ is $2 \times n_{c} \times N_{in}$ where $n_{c}$ is the number of orders; in this case $n_{c} = 2$, for NLO and N2LO.

The remaining piece in Eq.~(\ref{eq:marginalization}) is $P(\bar{c}^{2} | \boldsymbol{c}, \boldsymbol{\theta}, \Lambda_{B}, I)$. Since the coefficients are directly related to $\boldsymbol{\theta}$ and $\Lambda_{B}$, writing this piece of the conditional is redundant and the conditional pdf can be simplified to $P(\bar{c}^{2} | \boldsymbol{c}, I)$. This can then be computed using Bayes' theorem
\begin{equation}
	\label{eq:bayes_c}
	P(\bar{c}^{2} | \boldsymbol{c}, I) = \frac{P(\boldsymbol{c} | \bar{c}^{2}, I) P(\bar{c}^{2} | I)}{P(\boldsymbol{c} | I)}.
\end{equation}
Then, using Eq.~(\ref{eq:bar_c_prior}) along with the fact that $P(\boldsymbol{c} | I)$ normalizes the distribution we obtain
\begin{equation}
	\label{eq:c_bar_squared_prior}
	P(\bar{c}^{2} | \boldsymbol{\theta}, \Lambda_{B}, I) = \frac{{\left( \nu \tau^{2} / 2 \right)}^{\frac{\nu}{2}}}{\Gamma(\frac{\nu}{2}) {(\bar{c}^{2})}^{1 + \frac{\nu}{2}}} \exp \left( \frac{- \nu \tau^{2}}{2 \bar{c}^{2}} \right).
\end{equation}
By enforcing the conjugacy of the prior, we can update the hyperparameters $\nu$ and $\tau^{2}$ based on the order-by-order expansion coefficients according to
\begin{align}
	\label{eq:hyperparameter_update}
	\nu & = 2 n_c N_{in} + \nu_{0}, \\
	\nu \tau^{2} & = \nu_{0} \tau_{0}^{2} + \boldsymbol{c}^{2}. \nonumber 
\end{align}
In the limit as the number of degrees of freedom gets large, the peak of this distribution approaches the mean square value of the coefficients.

Lastly, we can focus on the second term in the joint intermediate-prior distribution of Eq.~(\ref{eq:intermediate_prior}), $P(\Lambda_{B} | \boldsymbol{\theta}, I)$. To start, we fist consider the distribution of the contribution to the partial wave scattering amplitude at order $n$, $P(\Delta \boldsymbol{f}| \Lambda_{B}, I)$. The elements of $\Delta \boldsymbol{f}$ are related to the elements of $\boldsymbol{c}$ by $\Delta f_{c, n}^{(\ell)}(k_j) = f_{c, 0}^{(\ell)}(k_j) c_{c, n, j}^{(\ell)} \left( \frac{k_j}{\Lambda_{B}} \right)^{n}$. Note that the $\Delta$ is here to distinguish this from the cumulative amplitude up to order $n$, defined in Eq.~(\ref{eq:f_order}). $\Delta \boldsymbol{f}$ is therefore 
implicitly dependent on $\boldsymbol{\theta}$ because the $f_{c, n}^{(\ell)}(k_j)$ are partial model-evaluations using the physics model parameters at $E_{j}$. Therefore we can convert the pdf of $\Delta \boldsymbol{f}$ to a pdf for $\boldsymbol{\theta}$ using a Jacobian. 
Performing the change of variables yields:
\begin{align}
	P(\boldsymbol{\theta} | \Lambda_{B}, I) =  P(\boldsymbol{c} | \Lambda_{B}, I) \frac{d \boldsymbol{c} }{d \Delta \boldsymbol{f}} \frac{d \Delta \boldsymbol{f}}{d \boldsymbol{\theta}} \nonumber \\
	= \prod_{n, \ell, j} \frac{P(c_{c, n, j}^{(\ell)} | \Lambda_{B}, I)}{\big| f_{c, 0}^{(\ell)} (k_j) \left( \frac{k}{\Lambda_{B}} \right)^{n} \big|} \frac{d \Delta f_{c, n}^{(\ell)}(k_j)}{d \boldsymbol{\theta}}.
\end{align}
Applying Bayes' theorem to the left-hand side of this expression lets us write
\begin{flalign}
	\label{eq:int_equation}
	P(\Lambda_{B} | \boldsymbol{\theta}, I) = \prod_{n, \ell, j} \frac{P(c_{c, n, j}^{(\ell)} | \Lambda_{B}, I) P(\Lambda_{B} | I)}{P(\boldsymbol{\theta} | I) \big| f_{c, 0, j}^{(\ell)} \left( \frac{k}{\Lambda_{B}} \right)^{n} \big|} \frac{d \Delta f_{c, n, j}^{(\ell)}}{d \boldsymbol{\theta}}.&&
\end{flalign}
Now we follow Refs.~\cite{Melendez:2017,Melendez:2019} and compute  $P(c_{c, n, j}^{(\ell)} | \Lambda_{B}, I)$ by using Eq.~(\ref{eq:bar_c_prior}) in Eq.~(\ref{eq:bayes_c}), and then equating the result with Eq.~(\ref{eq:c_bar_squared_prior}). This yields an explicit form for $P(\boldsymbol{c}|\Lambda_B,I)$. We can then drop all factors in (\ref{eq:int_equation})
 that do not depend on $\Lambda_{B}$ to obtain the unnormalized intermediate-prior
\begin{equation}
	P(\Lambda_{B} | \boldsymbol{\theta}, I) \propto \frac{P(\Lambda_{B} | I)}{\tau^{\nu} \prod_{n, \ell, j} \big| \Lambda_{B}^{-n} \big|}.
\end{equation}
Now we do the product over the two different partial waves, and also over the number of data points $N_{in}$, giving us
\begin{equation}
	\label{eq:Lambda_B_prior}
	P(\Lambda_{B} | \boldsymbol{\theta}, I) \propto \frac{P(\Lambda_{B} | I)}{\tau^{\nu} \prod_{n} \big| \Lambda_{B}^{-n} \big|^{2 N_{in}}}.
\end{equation}
In order to normalize this distribution at a specific $\boldsymbol{\theta}$, we perform a numerical integration over the range of $\Lambda_{B} \in [\max\{k\}, 6 \text{ fm}^{-1}]$. This normalization is completed at every MCMC step. The lower limit is determined based on the condition that $Q < 1$ for the EFT expansion to be valid, and the upper limit is chosen to be sufficiently large such that the prior is not sensitive to this cutoff. 

Lastly we choose the prior on the breakdown scale to be
\begin{equation}
	\label{eq:Lambda_B_prior_choice}
	P(\Lambda_{B} | I) \sim \mathcal{N} \left( 1.0 \text{ fm}^{-1}, \left[0.7 \text{ fm}^{-1}\right]^{2} \right).
\end{equation}

Now that we have constructed every piece of the joint posterior probability distribution, we can sample from it using various Monte Carlo sampling techniques. 

%% file: results_no_f_waves.tex
\section{Implementation and Results Without $f$-Waves}
\label{sec:implementation_without_f_waves}
In this section, we present the results of simultaneous sampling of the model parameters and the EFT truncation uncertainty parameters for analysis of part of the Paneru dataset using a scattering model without $f$-waves. We perform two different analyses: one with $E_{\text{max}} = 2.6$ MeV ($7$ energy bins) and another with $E_{\text{max}} = 3.6$ MeV ($8$ energy bins). We modify the theory covariance matrix $\Sigma^{\text{th}}$ from the one defined using Eqs.~(\ref{eq:delta_sigma}) \& (\ref{eq:theory_covariance}) to account for the fact that we are not including the $f$-waves. This is done by setting $f_{c}^{(3)} = 0$ and $f_{i}^{(3)} = 0$ in Eq.~(\ref{eq:delta_sigma}). For each analysis, we sample the normalization for each energy bin ($\bm{\xi}$) along with $\boldsymbol{a}$ as defined in Eq.~(\ref{eq:a_params}) and $\{\bar{c}^{2}, \Lambda_{B} \}$.  

In this analysis, we used the \texttt{ptemcee} implementation of the parallel tempered Monte Carlo sampling method~\cite{Vousden_2015}. This method of sampling is a modification to the better known Metropolis-Hastings algorithm which allows for a more efficient sampling of the space. This parallel tempered Monte Carlo method extends the Metropolis-Hastings algorithm by introducing a ``temperature ladder''. This temperature ladder, $\beta \in (0, 1]$ acts as a scaling factor for the likelihood function. Here we set up a ladder of $8$ temperatures with $\beta = \{1.0, 0.92, 0.84, 0.77, 0.71, 0.5, 0.35\}$. In the traditional Metropolis-Hastings algorithm, the sampler proposes the next step in parameter space, but in parallel tempered Monte Carlo, the sampler can also propose to swap the current state with a state from a different temperature. This allows the sampler to explore the parameter space more efficiently and typically produces very low autocorrelation lengths within sample chains.

We initialize the sampler at each temperature with a set of walkers, each starting from a random point according to the prior distributions. The number of walkers is set to be twice the total number of parameters being sampled, i.e., $26$ $(28)$ for $E_{max} = 2.6$ $(3.6)$ MeV. The sampler is then run for a total of $12,000$ steps, with the first $4,000$ steps being discarded as burn-in. The remaining $8,000$ steps are then thinned according to the autocorrelation time of the chain (with \texttt{ptemcee} this is typically $\approx 2$) to produce independent samples. After thinning we are left with approximately $4,000$ independent samples for each walker and each temperature. The final posterior is obtained by combining all the samples with temperature $\beta = 1.0$.

\begin{figure*}[ht]
    \centering
    \includegraphics[width = 0.9\textwidth]{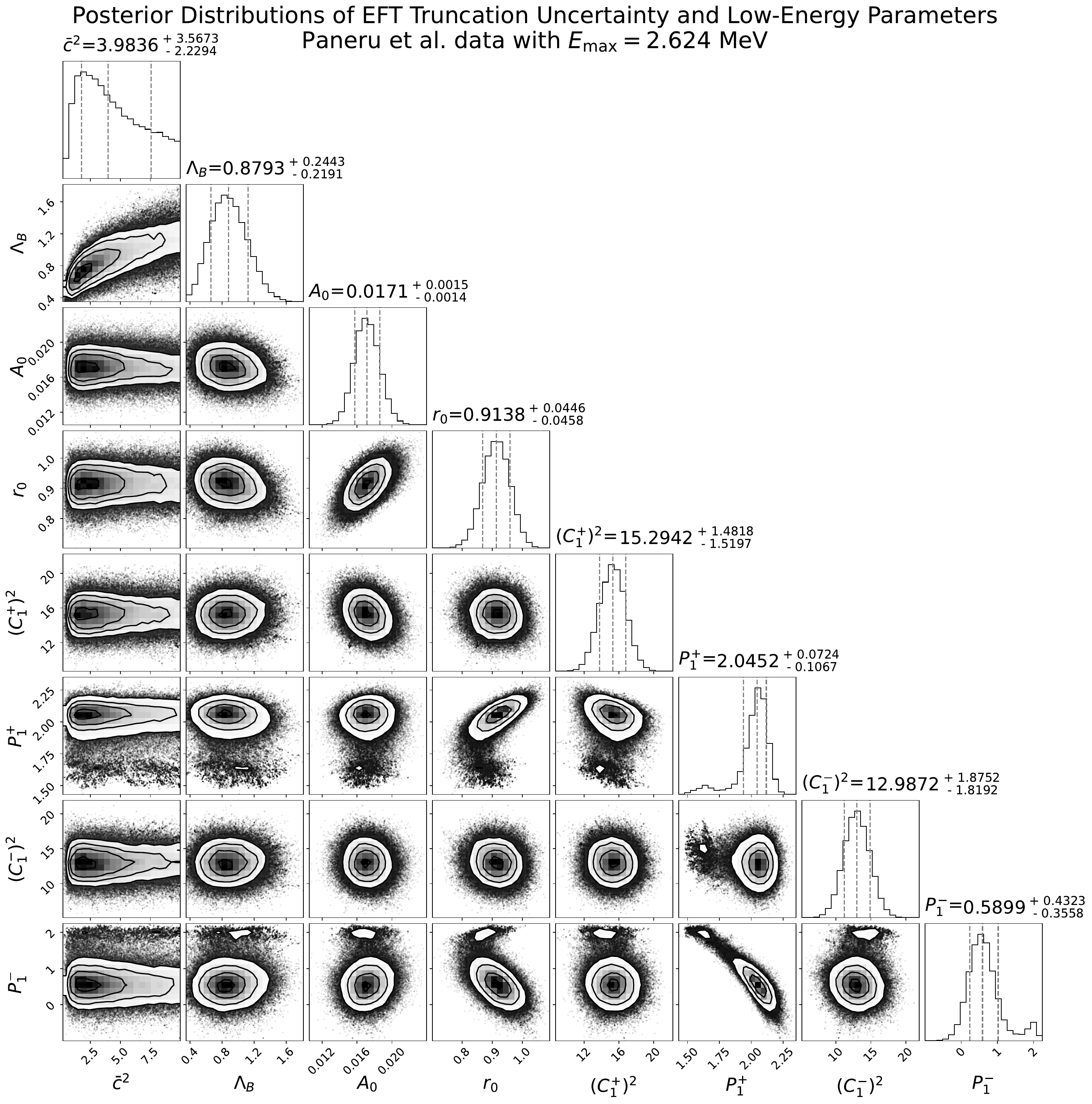}
    \caption{\label{fig:no_f_2-6_corner}Corner plot of the joint posterior distribution of the model parameters and the truncation uncertainty parameters for analysis without the $f$-waves up to $E_{\text{max}} = 2.6$ MeV.}
\end{figure*}

The results from the analysis without $f$-wave interactions with $E_{\text{max}} = 2.6$ MeV are shown in Fig.~\ref{fig:no_f_2-6_corner}. The corner plot shows the joint posterior distribution of the model parameters and the truncation uncertainty parameters. 

Considering all data up to $E_{\text{max}} = 2.6$ MeV from the Paneru dataset, we infer an EFT breakdown scale of $\Lambda_{B} = 0.88 ^{+0.24}_{-0.22} \text{ fm}^{-1}$. This result is consistent with the EFT breakdown scale estimated by Poudel \& Phillips~\cite{Poudel_2022}. Poudel \& Phillips estimated a breakdown scale of $1 \text{ fm}^{-1}$ based on the sizes of the $^{4}$He and $^{3}$He nuclei. The inferred value of $\bar{c}^{2} = 3.98 ^{+3.57}_{-2.23}$ is consistent with the expectation that the expansion coefficients are $\mathcal{O}(1)$ numbers. The estimated ERPs and ANCs from this analysis are also consistent with previous similar analyses~\cite{Burnelis:2024,Poudel_2022}. 

We also did an analysis without the $f$-wave interactions up to $E_{\text{max}} = 3.6$ MeV. The parameter distributions from this analysis are not consistent with any previous analyses. This is likely due to the fact that the $f$-wave effects are starting to enter the cross section. While we may not have enough data in this energy range to accurately constrain the $f$-wave parameters, we start to see that the $s$ and $p$-wave parameters change dramatically compared to the $E_{max} = 2.6$ MeV analysis to account for the missing $f$-wave contributions. 

The simultaneous sampling analysis up to $E_{\text{max}} = 3.6$ MeV yields a breakdown scale of $\Lambda_{B} = 0.72^{+0.17}_{-0.16} \text{ fm}^{-1}$. This value being lower, in combination with a larger estimate for $\bar{c}^{2}$ shows that the model is trying to compensate for the missing $f$-wave contributions by increasing the model uncertainty. It is worth noting that extending the analysis to $E_{\text{max}} = 4.4$ MeV further exacerbates this issue, with the inferred breakdown scale decreasing even more and the model parameters becoming even less consistent with previous analyses. The breakdown scale estimated from $E_{\text{max}} = 4.4$ MeV is near the resonance momentum, and the expansion coefficients are no longer $\mathcal{O}(1)$ numbers.

\begin{figure}
    \centering
    \includegraphics[width = 0.9\columnwidth]{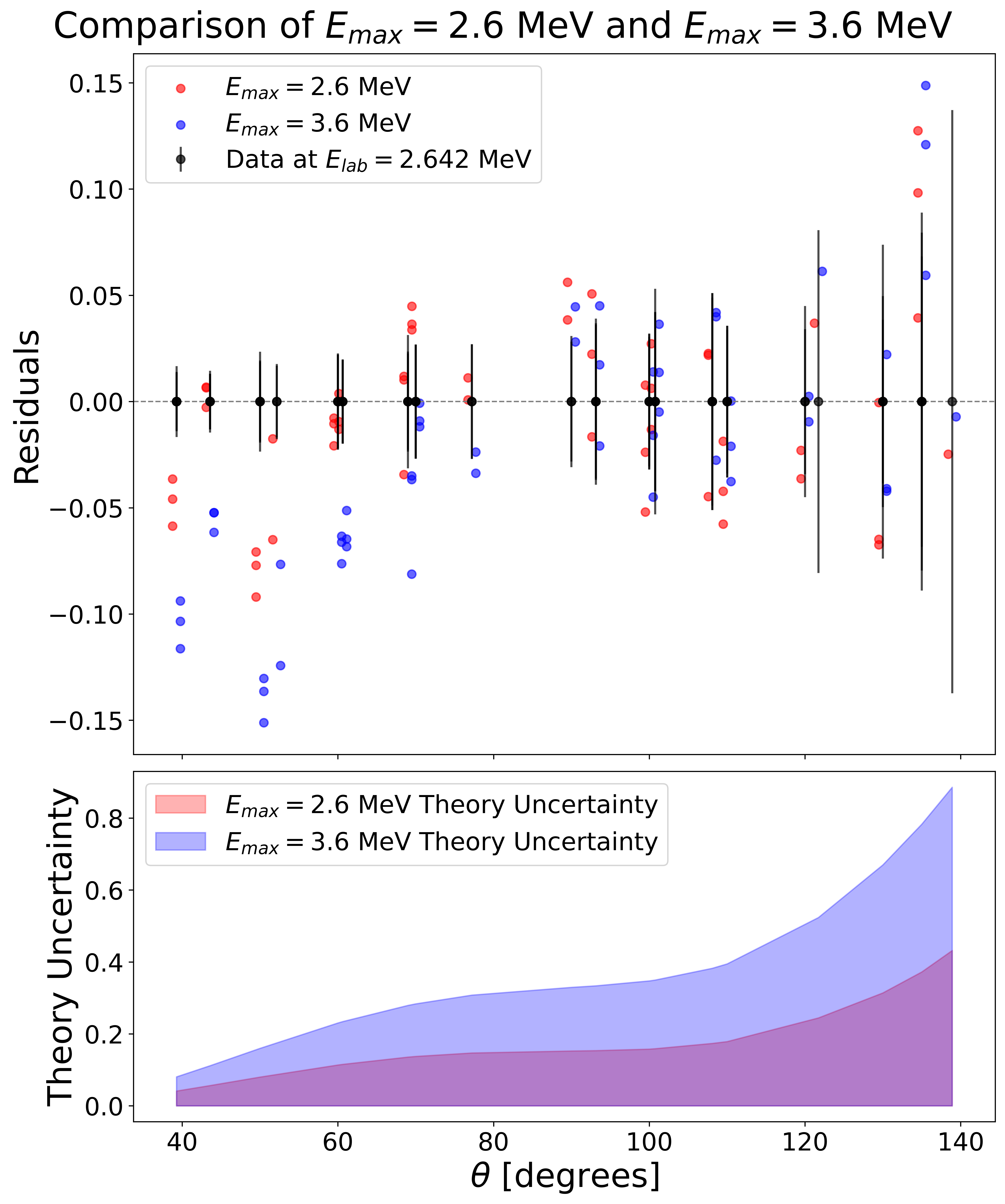}
    \caption{\label{fig:res_plot2-3} Top panel: Residuals of the data with respect to the model prediction for analysis without the $f$-wave interactions at $E_{\text{max}} = 2.6$ MeV (red points) and $E_{\text{max}} = 3.6$ MeV (blue points). The vertical lines indicate the experimental uncertainties. Bottom panel: Model uncertainty (upper-half) envelope for the same analysis. The colors correspond to the same analyses as in the top panel.}
\end{figure}

A comparison of the residuals for the two models fit up to $E_{\text{max}} = 2.6$ MeV and $E_{\text{max}} = 3.6$ MeV is shown in Fig.~\ref{fig:res_plot2-3}. The residuals for the model fit up to $E_{\text{max}} = 3.6$ MeV show a larger systematic trend, indicating that the model is not accurately capturing the data in the fitting region. Additionally, the model uncertainty envelope is nearly twice as large for the $E_{\text{max}} = 3.6$ MeV analysis compared to the $E_{\text{max}} = 2.6$ MeV analysis, providing more evidence that the model is struggling to account for the missing $f$-wave contributions. All residuals, except for the evaluation at $40^{\circ}$ for $E_{\text{max}} = 3.6$ MeV, are within $1 \sigma$ of the theoretical model uncertainty for both analyses.

\begin{figure*}
    \centering
    \includegraphics[width = 0.9\textwidth]{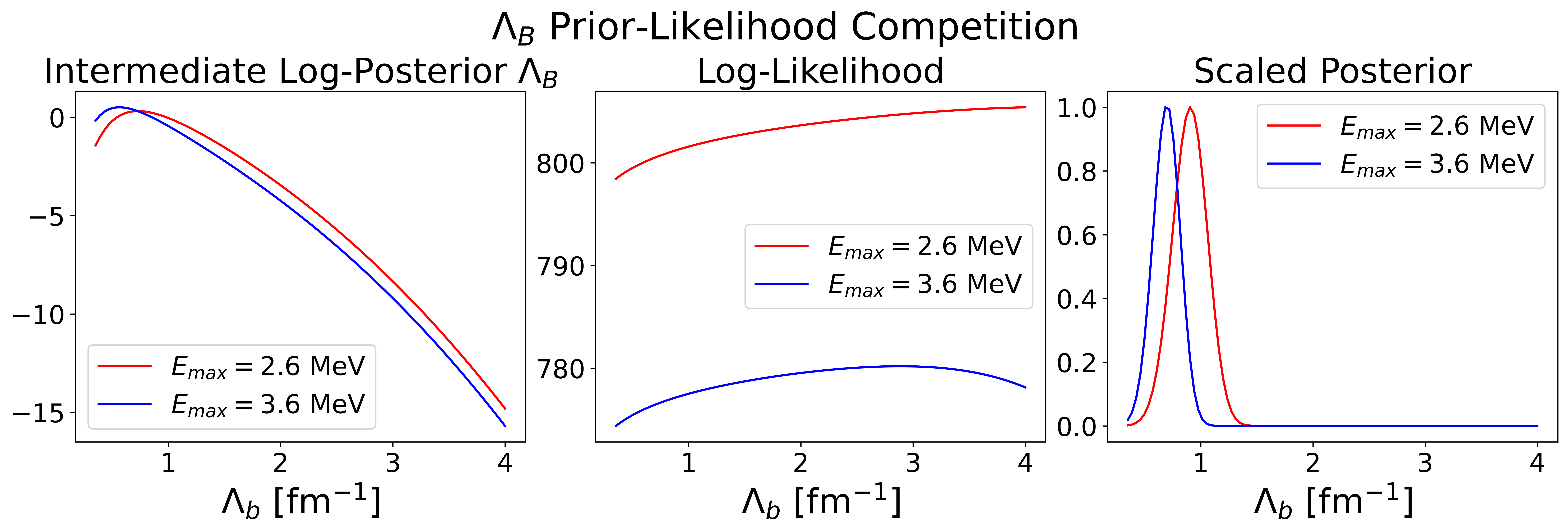}
    \caption{\label{fig:competition_plot2-3} Competition between the likelihood and the intermediate-prior distribution of $\Lambda_{B}$ for the analysis without the $f$-waves up to $E_{\text{max}} = 2.6$ MeV (red curve) and $E_{\text{max}} = 3.6$ MeV (blue curve). The left panel corresponds to the log of Eq.~(\ref{eq:likelihood}), the middle panel corresponds to the log of Eq.~(\ref{eq:Lambda_B_prior}), and the right panel corresponds to the full posterior distribution of $\Lambda_{B}$. We scale the posterior distribution such that the maximum value is $1$.}
\end{figure*}

Figure~\ref{fig:competition_plot2-3} shows the competition between the log-likelihood (log of Eq.~(\ref{eq:likelihood})) and the log-intermediate prior distribution of $\Lambda_{B}$ (log of Eq.~(\ref{eq:Lambda_B_prior})). To generate this figure, we take the maximum a posteriori estimates of all the parameters except for $\Lambda_{B}$, and then compute the log-likelihood and log-intermediate prior distributions for a range of $\Lambda_{B}$ values. In the case of $E_{\text{max}} = 2.6$ MeV, the likelihood continues growing as $\Lambda_{B}$ increases and peaks at approximately $\Lambda_{B} = 6$ fm$^{-1}$, while the intermediate prior is peaked at approximately $\Lambda_{B} = 0.75$ fm$^{-1}$. The combination of these two distributions yields a posterior distribution that is peaked at approximately $1$ fm$^{-1}$. In the analysis with $E_{\text{max}} = 3.6$ MeV, the likelihood is peaked at approximately $3$ fm$^{-1}$, while the intermediate prior is peaked at approximately $0.5$ fm$^{-1}$. The posterior distribution is then peaked at approximately $0.7$ fm$^{-1}$.

%% file: results_f_waves.tex
\section{Implementation and Results Including $f$-Waves}
\label{sec:implementation_with_f_waves}
Here we present the results of the simultaneous sampling of the model parameters and the EFT truncation uncertainty parameters for the analysis of the Paneru dataset including the full $f$-wave treatment. We perform the analysis with $E_{\text{max}} = 5.5$ MeV (including all the data in the set). For this analysis, we sample the set of $10$ normalization parameters and augment the set of parameters $\boldsymbol{a}$ to include the $f$-wave parameters $r_{3}^{+}$ and $P_{3}^{+}$. 

We use the same prior distributions as defined in Eqs.~(\ref{eq:a_prior}), (\ref{eq:c_bar_squared_prior}), and (\ref{eq:Lambda_B_prior_choice}). The model parameter prior distributions are defined in Tab.~\ref{tab:param_priors1}. The normalization prior parameters are defined in Tab.~\ref{tab:norm_priors}. We use the fully constructed theory covariance matrix defined in Eq.~(\ref{eq:theory_covariance}) to model the theory uncertainty. We again follow the same procedure as described in Sec.~\ref{sec:implementation_without_f_waves} and sample the joint posterior distribution.

\begin{figure*}
    \centering
    \includegraphics[width = 0.95\textwidth]{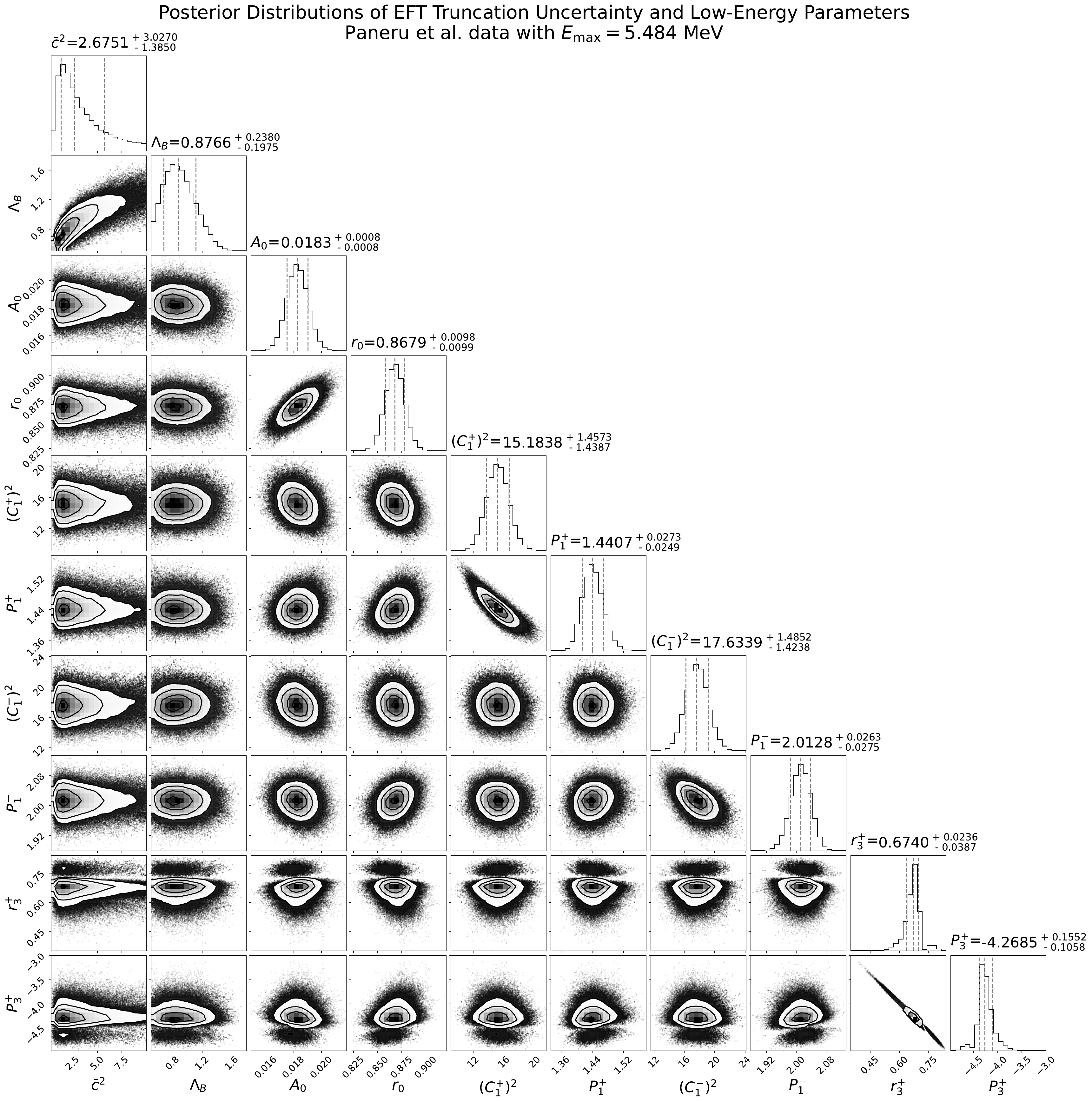}
    \caption{\label{fig:full_f_wave_corner} Corner plot of the joint posterior distribution of the model parameters and the truncation uncertainty parameters for analysis including the $\frac{7}{2}^{-}$ and $\frac{5}{2}^{-}$ $f$-wave channels up to $E_{\text{max}} = 5.5$ MeV.}
\end{figure*}

The joint posterior distribution of the model parameters and the truncation uncertainty parameters is shown in Fig.~\ref{fig:full_f_wave_corner}. The posterior distribution of the $f$-wave effective range parameters $r_{3}^{+}$ and $P_{3}^{+}$ shows a bimodal structure. The two modes correspond to two different solutions for the $\frac{7}{2}^{-}$ phase shifts. The highly anti-correlated structure of these parameters is expected due to the fact that these are competing parameters in the condition from Eq.~(\ref{eq:resonance_condition}). 

This analysis provides a large and positive $s$-wave scattering length $a_{0} = 54.64$ fm$^{-1}$. The $s$-wave effective range parameter is positive and of natural size $r_{0} = 0.87$ fm. The $p$-wave ANC for the $1^{+}$ channel is stable and well constrained, while the other $p$-wave parameters have significant migration from where they were in the analysis without $f$-wave interactions. Both $p$-wave shape parameters $P_{1}^{\pm}$ have moved and are much better constrained.

\begin{figure*}
    \centering
    \includegraphics[width = 0.9\textwidth]{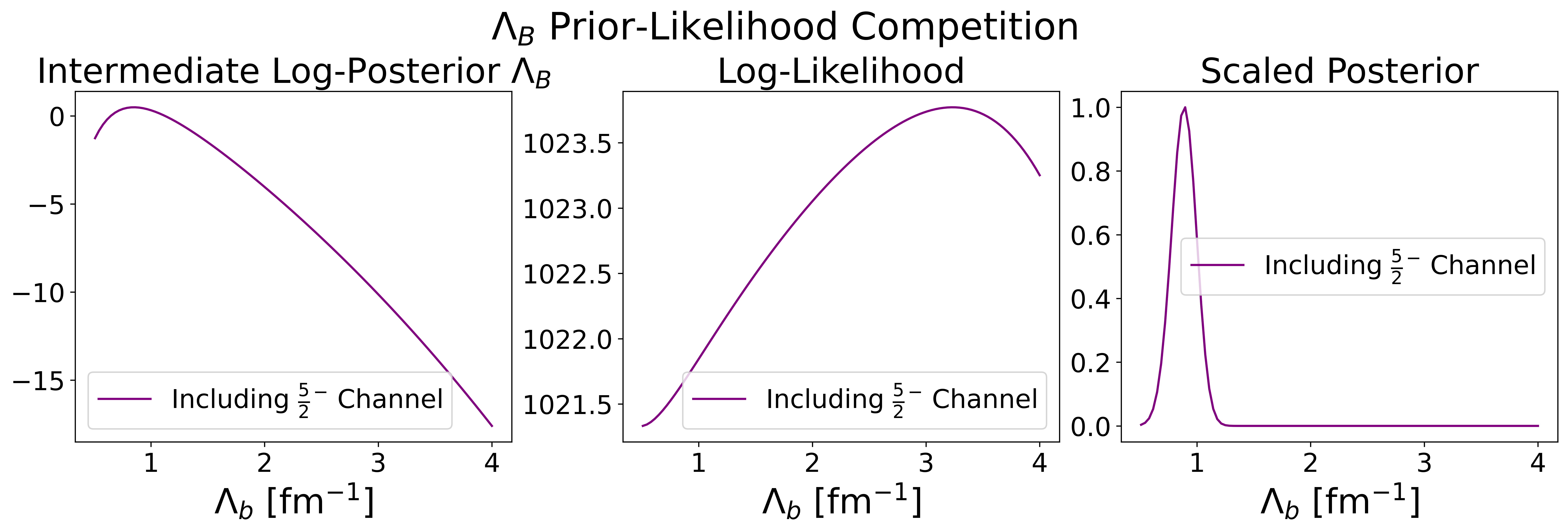}
    \caption{\label{fig:competition_plot5-5} Competition between the likelihood and the intermediate-prior distribution of $\Lambda_{B}$ for the analysis including the $f$-waves up to $E_{\text{max}} = 5.5$ MeV. The left panel corresponds to the log of Eq.~(\ref{eq:likelihood}), the middle panel corresponds to the log of Eq.~(\ref{eq:Lambda_B_prior}), and the right panel corresponds to the full posterior distribution of $\Lambda_{B}$. We scale the posterior distribution such that the maximum value is $1$.}
\end{figure*}

The estimated EFT breakdown scale is $\Lambda_{B} = 0.90^{+0.24}_{-0.21} \text{ fm}^{-1}$, consistent with Poudel \& Phillips~\cite{Poudel_2022}. This value is close to the estimated breakdown scale from the analysis without $f$-waves up to $E_{\text{max}} = 2.6$ MeV. Figure~\ref{fig:competition_plot5-5} shows the competition between the likelihood and the intermediate prior distribution of $\Lambda_{B}$. This figure shows the multiple components that contribute to the posterior distribution of $\Lambda_{B}$. We see that the intermediate prior is peaked at approximately $0.86$ fm$^{-1}$, while the likelihood is peaked at approximately $3.2$ fm$^{-1}$. The likelihood in this case gently pulls the posterior up just a bit to $0.90$ fm$^{-1}$. The estimated value of $\bar{c}^{2} = 2.68^{+3.03}_{-1.39}$, indicating that the expansion coefficients are natural in size. 

\begin{figure}[h!]
    \centering
    \includegraphics[width = 0.95\columnwidth]{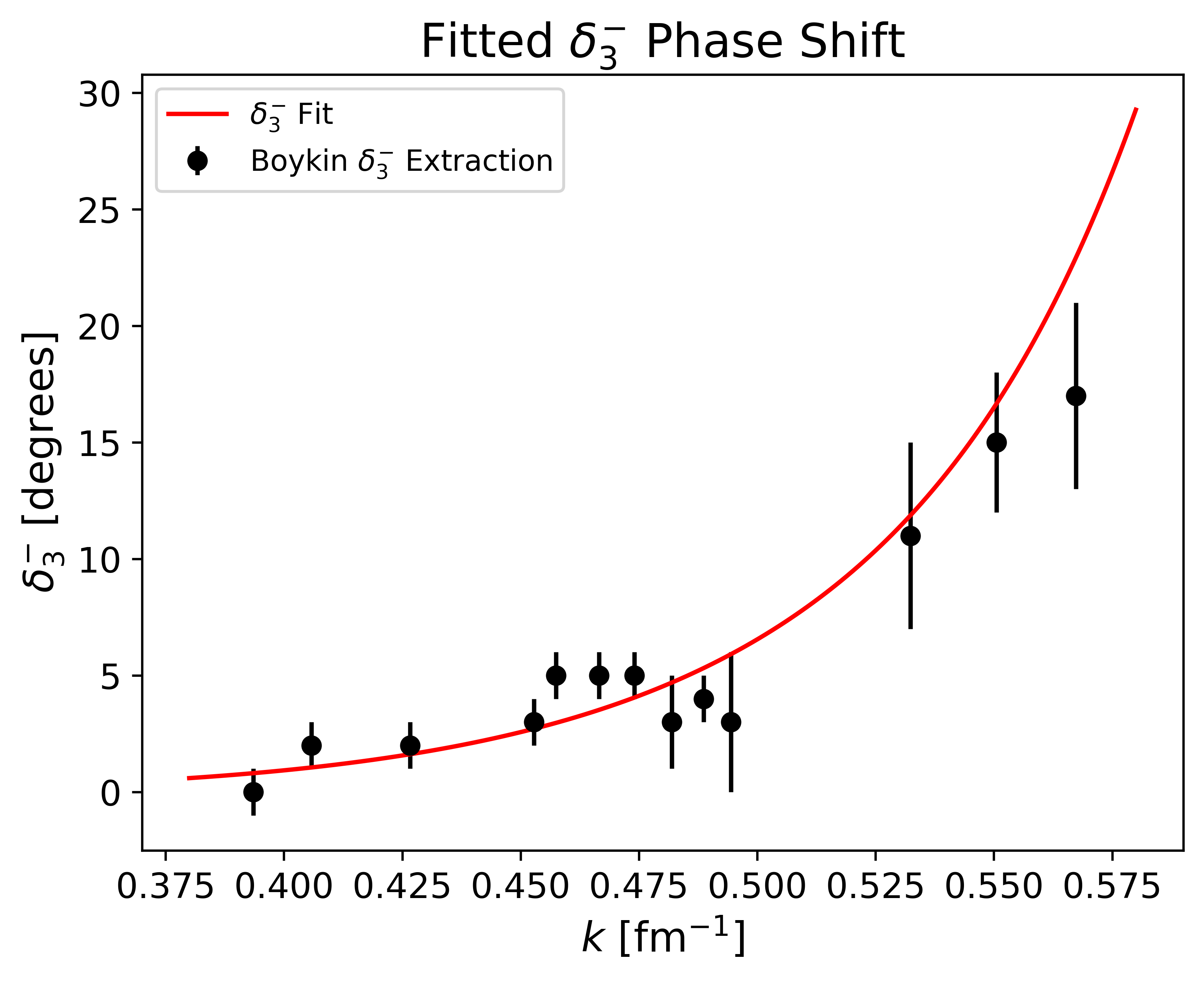}
    \caption{\label{fig:boykin_phase_shifts}Boykin phase shift extraction for the $\frac{5}{2}^{-}$ channel and the best fit curve. The best fit values are $a_{3}^{-} = -108.28$ fm$^{7}$ and $r_{3}^{-} = 0.07$ fm$^{-5}$, and the $\chi^{2} / \text{dof} = 1.24$.}
\end{figure}

When including the $f$-waves into the scattering model, we also include the $\frac{5}{2}^{-}$ channel. This channel has a broad resonance at $E = 6.73$ MeV~\cite{Tilley2002}. This resonance is high enough in energy that the Paneru data set cannot accurately constrain the $\frac{5}{2}^{-}$ channel's ERPs. Thus, we do not sample to infer the ERPs of this channel, but instead we fit the effective range $r_{3}^{-}$ and tune the scattering length $a_{3}^{-}$ to both reproduce the Boykin phase shifts and the resonance at $E = 6.73$ MeV. The best fit values are $a_{3}^{-} = -108.28$ fm$^{7}$ and $r_{3}^{-} = 0.07$ fm$^{-5}$. These values yield a $\chi^{2} / \text{dof} = 1.24$ when compared to the Boykin $\delta_{3}^{-}$ phase shifts. The Boykin phase shift extractions and the fit are shown in Fig.~\ref{fig:boykin_phase_shifts}.

\begin{figure*}
    \centering
    \includegraphics[width = 0.95\textwidth]{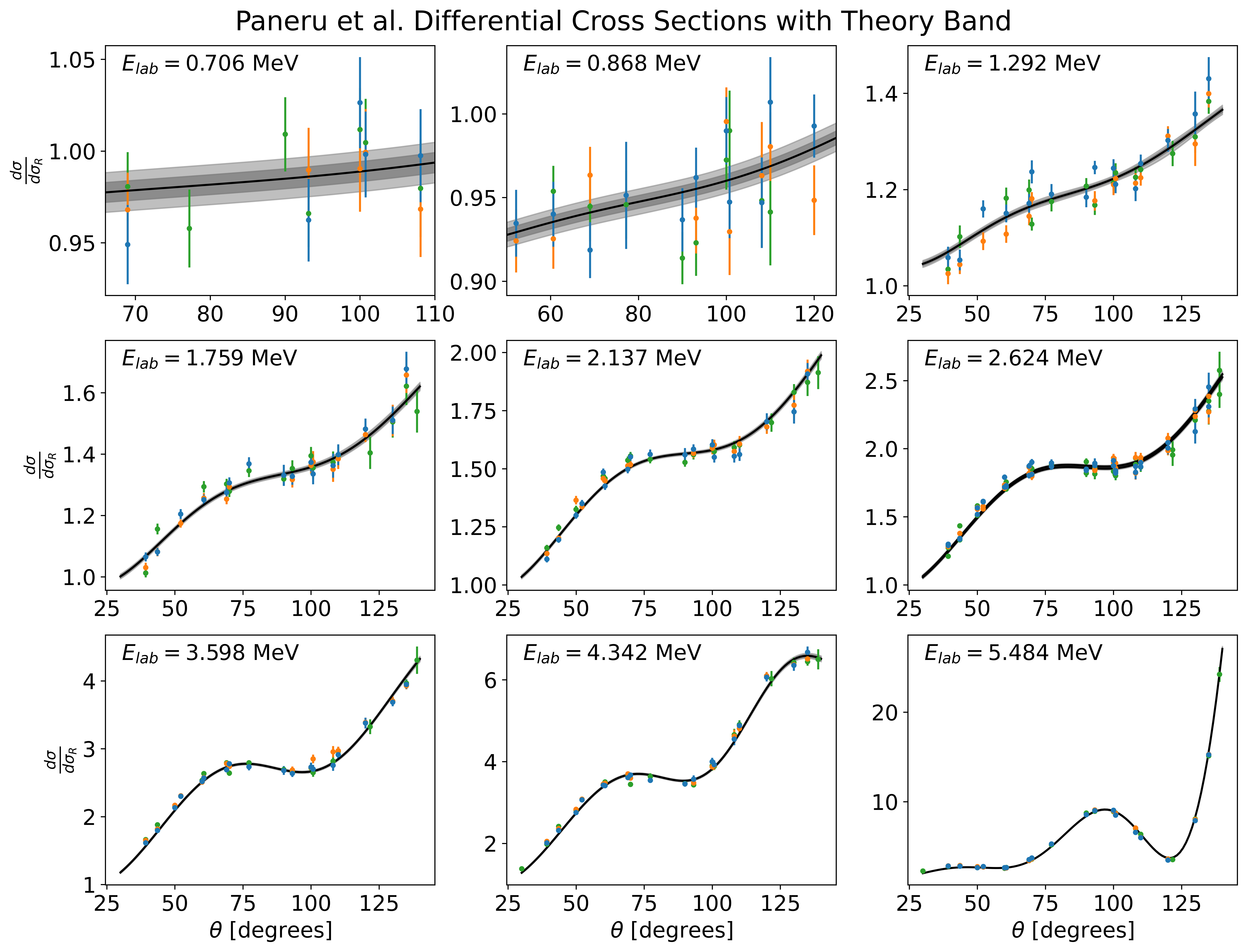}
    \caption{\label{fig:cs_theory} Differential cross section predictions (relative to Rutherford) for the analysis including the full $f$-wave interaction. The dark and light shaded regions represent the $68\%$ and $95\%$ theory uncertainty bands, respectively.}
\end{figure*}

We show the differential cross section predictions for the analysis including the full $f$-wave interaction in Fig.~\ref{fig:cs_theory}. The uncertainty bands represent the $68\%$ and $95\%$ theory uncertainty envelopes. To compute these bands, we overlay a large representative sample of the posterior predictive distributions and compute the percentiles at each angle. The predictions accurately describe the data across the entire energy range: the residuals for this model that includes the $\frac{5}{2}^{-}$ channel are within the theory uncertainty envelope, indicating that we are accurately describing the data within the model uncertainty.

\begin{figure}[h!]
    \centering
    \includegraphics[width = 0.95\columnwidth]{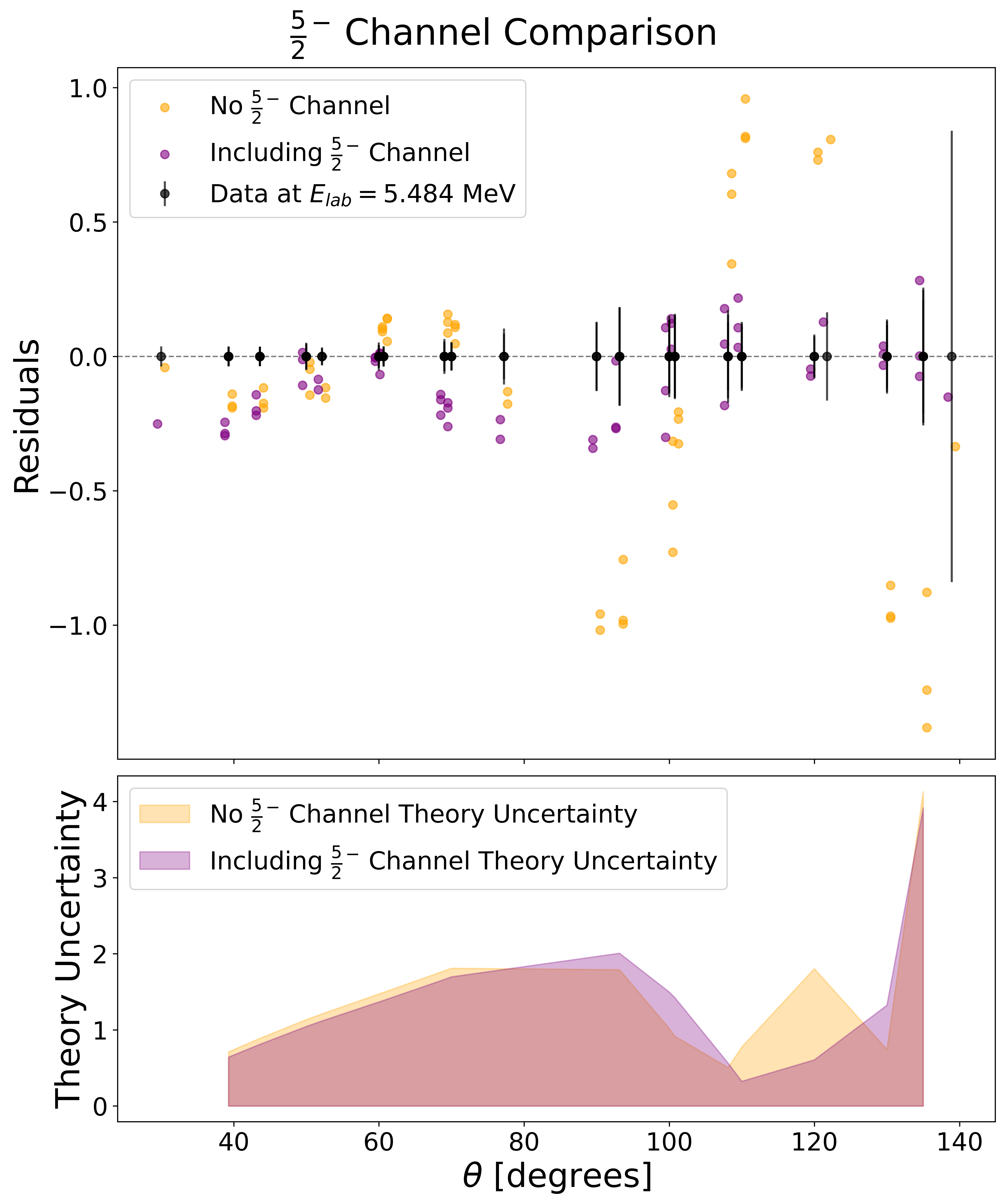}
    \caption{\label{fig:f_wave_residuals}Top panel: Residuals of the Paneru cross section measurements (relative to Rutherford) at $E = 5.5$ MeV with respect to the model prediction for two analyses: one that adds just the $\frac{7}{2}^-$ channel to the $s$- and $p$-wave amplitude, and one that includes both $f$-waves channels. Bottom panel: The shaded region represents the size of the model uncertainty envelope for the same two analyses.}
\end{figure}

Finally, we point out that the $\frac{5}{2}^{-}$ channel's contribution to the cross section is important in this full $f$-wave treatment. Fig.~\ref{fig:f_wave_residuals} shows the comparison of residuals for models that include and exclude this channel. The residuals for the model that does not include the $\frac{5}{2}^{-}$ channel are larger and show a systematic trend across the angular range. 

%% file: conclusion.tex
\section{Conclusion}
\label{sec:conclusion}
Extending the effective field theory framework to include the $f$-wave channels provides a more complete description of the $^{3}$He-$\alpha$ system. In this work, we have presented a Bayesian analysis of the low-energy elastic scattering of $^{3}$He and $^{4}$He that includes the $f$-wave channels and simultaneously estimates the EFT model uncertainty parameters. The $\frac{7}{2}^{-}$ resonance is included explicitly in the scattering model, and the $\frac{5}{2}^{-}$ channel is included  
by fitting the $\frac{5}{2}^{-}$ channel's effective range and scattering length to the Boykin phase shifts~\cite{Boykin:1972}. 

To accurately model the uncertainty in the scattering model, we have developed a sophisticated theory covariance matrix that operates at the amplitude level. This covariance matrix captures the angular dependence of the cross section and allows us to define the model uncertainty in two different regions: below the resonance window, and within the resonance window. 

Simultaneously sampling the model parameters and the EFT truncation uncertainty parameters allows us to estimate the EFT breakdown scale $\Lambda_{B}$. By comparing two reasonable analyses, one at $E_{\text{max}} = 2.6$ MeV without the $f$-waves, and one at $E_{\text{max}} = 5.5$ MeV with the $f$-waves, we see that the estimated breakdown scale is stable with estimated values of  $\Lambda_{B} = 0.88^{+0.24}_{-0.22} \text{ fm}^{-1}$ and $\Lambda_{B} = 0.90^{+0.24}_{-0.21} \text{ fm}^{-1}$. 

We make an assumption here that the EFT breakdown scale $\Lambda_{B}$ is the same breakdown scale for all channels. This assumption may not be true because by our treatment of the $\frac{7}{2}^{-}$ resonance, we are performing a different expansion in the different kinematic regions. Our expansion below the resonance window is an expansion about how far in momentum we are from $k = 0$, while within the resonance window the expansion is about how far away we are from the resonance momentum $k = k_{R}$. For this reason, we suspect that there may be two different breakdown scales, one for the low-energy region and one for the vicinity near the resonance. If this is the case, then the EFT framework presented here can be extended to include this additional breakdown scale.

Some potential future work would be to include the Barnard scattering data~\cite{Barnard:1964}, as well as $^{3}$He$(\alpha, \gamma)$ capture data to better constrain the ERPs. Including this data would allow us to make a more precise prediction for the ANCs and the astrophysical $S$-factor $S_{34}(0)$. The methodology and the framework presented here can also be applied to other reactions of interest. A natural next step would be to apply this framework to the $^{7}$Be + $p$ reaction, the next step in the $pp$-II chain.

%% file: width.tex
\section{Resonance Derivations}
\label{app:width}
When considering a resonance in a specific partial wave, it is important to relate the resonance parameters to the relevant ERPs in that channel. In this appendix, we derive the relationships in Eqs.~(\ref{eq:resonance_condition})--(\ref{eq:tau_3}) that relate the $f$-wave ERPs to the resonance position and width. We begin with the expression for the $\frac{7}{2}^{-}$ inverse scattering amplitude:
\begin{align}
    \label{eq:f_invamp}
    \frac{1}{k^{7} \left( \cot \delta_{3}^{+} - i \right)} = C_{0}^{2} V_{3}(\eta) \bigg[ 36 \bigg\{ -\frac{1}{a_{3}^{+}} + \frac{r_{3}^{+}}{2} k^{2} \nonumber \\
    + \frac{P_{3}^{+}}{4} k^{4} \bigg\} - 2 k_{c} k^{6} V_{3}(\eta) H(\eta)\bigg]^{-1}.
\end{align}
At the momentum corresponding to the resonance pole $k_{R}$, the real part of the quantity in square brackets in Eq.~(\ref{eq:f_invamp}) is zero. This condition allows us to directly tune the $f$-wave scattering length $\frac{1}{a_{3}^{+}}$ giving us
\begin{align}
    \label{eq:f_resonance_condition}
    \frac{1}{a_{3}^{+}} = \frac{r_{3}^{+}}{2} k_{R}^{2} + \frac{P_{3}^{+}}{4} k_{R}^{4} - \Bigg( \frac{2 k_{c} k_{R}^{6}}{36}  (1 + \eta_{R}^{2}) \nonumber \\ \times (4 + \eta_{R}^{2}) (9 + \eta_{R}^{2}) \Re \left[ H(\eta_{R}) \right] \Bigg).
\end{align}

It is useful to define the function
\begin{align}
    \label{eq:denom}
    D(k) = 36 \bigg\{ -\frac{1}{a_{3}^{+}} + \frac{r_{3}^{+}}{2} k^{2} + \frac{P_{3}^{+}}{4} k^{4} \bigg\} \nonumber \\
    - 2 k_{c} k^{6} V_{3}(\eta) \Re \left[H(\eta)\right]
\end{align}
which is the real part of the denominator of the right-hand side of Eq.~(\ref{eq:f_invamp}).

We now expand $D(k)$ around the resonance momentum up to first order in $(k^2 - k^2_{R})$:
\begin{align}
    \label{eq:denom_expansion}
    D(k) = D(k_{R}) + \frac{d D(k)}{d k^{2}} \Bigg|_{k = k_{R}} (k^2 - k^2_{R}) \nonumber \\
    + \mathcal{O}[(k^2-k_R^2)^2].
\end{align}
By design, the first term in the expansion $D(k_{R}) = 0$ via tuning of the $f$-wave scattering length. Thus, we have the approximation
\begin{align}
    \label{eq:denom_approx}
    D(k) \approx \frac{d D(k)}{d k^{2}} \Bigg|_{k = k_{R}} (k^2 - k_{R}^2).
\end{align}
The derivative on the right-hand side is given by 
\begin{align}
    \label{eq:denom_derivative}
    \frac{d D(k)}{d k^{2}} & \Bigg|_{k = k_{R}} = 36 \left( \frac{r_{3}^{+}}{2} + \frac{P_{3}^{+}}{2} k_{R}^{2} \right) - \frac{d}{d k^{2}} \Bigg[ 2 k_{c} k^{6} (1 + \eta^{2}) \nonumber \\ 
    & \times (4 + \eta^{2}) (9 + \eta^{2}) \Re \left\{ H(\eta) \right\} \Bigg] \Bigg|_{k = k_{R}}.
\end{align}
We define this quantity as $\tau_{3}^{+}$. This yields a Breit-Wigner-like form of the scattering amplitude near the resonance pole:
\begin{equation}
    \label{eq:breit_wigner}
    \tan \delta_{3}^{+} = \frac{\Gamma_{R}(E)}{2}\frac{1}{E_{R} - E},
\end{equation}
where $E_{R}$ is the resonance energy, and $\Gamma_{R}(E)$ is the (energy-dependent) resonance width. We can take the real part of the inverse scattering amplitude in Eq.~(\ref{eq:f_invamp}) and use the approximation in Eq.~(\ref{eq:denom_approx}) to find
\begin{equation}
    \label{eq:real_invamp}
    \tan \delta_{3}^{+} = \frac{- k^{7} C_{3}^{2}}{2 \mu \tau_{3}^{+}}\frac{1}{(E - E_{R})}.
\end{equation}
Comparing Eqs.~(\ref{eq:breit_wigner}) and~(\ref{eq:real_invamp}), we identify the resonance width as 
\begin{equation}
    \label{eq:resonance_width}
    \Gamma_{R}(E) = \frac{ -k^{7} C_{3}^{2} }{ \mu \tau_{3}^{+} }.
\end{equation}

%% file: citations.bib
@article{Hammer:2017,
    author = "Hammer, H. -W. and Ji, C. and Phillips, D. R.",
    title = "{Effective field theory description of halo nuclei}",
    eprint = "1702.08605",
    archivePrefix = "arXiv",
    primaryClass = "nucl-th",
    doi = "10.1088/1361-6471/aa83db",
    journal = "J. Phys. G",
    volume = "44",
    number = "10",
    pages = "103002",
    year = "2017"
}

@article{Poudel_2022,
	doi = {10.1088/1361-6471/ac4da6},
	url = {https://doi.org/10.1088\%2F1361-6471\%2Fac4da6},
	year = 2022,
	month = {mar},
	publisher = {{IOP} Publishing},
	volume = {49},
	number = {4},
	pages = {045102},
	author = {Maheshwor Poudel and Daniel R Phillips},
	title = {Effective field theory analysis of $^{3}\uppercase{H}$e-$\alpha$ scattering data},
	journal = {Journal of Physics G: Nuclear and Particle Physics}
}

@article{Vousden_2015,
   title={Dynamic temperature selection for parallel tempering in Markov chain Monte Carlo simulations},
   volume={455},
   ISSN={1365-2966},
   url={http://dx.doi.org/10.1093/mnras/stv2422},
   DOI={10.1093/mnras/stv2422},
   number={2},
   journal={Monthly Notices of the Royal Astronomical Society},
   publisher={Oxford University Press (OUP)},
   author={Vousden, W. D. and Farr, W. M. and Mandel, I.},
   year={2015},
   month=nov, pages={1919–1937}
}

@article{deboer2014monte,
  title={Monte Carlo uncertainty of the He 3 ($\alpha$, $\gamma$) Be 7 reaction rate},
  author={DeBoer, RJ and G{\"o}rres, J and Smith, K and Uberseder, E and Wiescher, M and Kontos, A and Imbriani, Gianluca and Di Leva, Antonino and Strieder, F},
  journal={Physical Review C},
  volume={90},
  number={3},
  pages={035804},
  year={2014},
  publisher={APS}
}

@article{Adelberger2011,
  title = {Solar fusion cross sections. II. The $pp$ chain and CNO cycles},
  author = {Adelberger, E. G. and Garc\'{\i}a, A. and Robertson, R. G. Hamish and Snover, K. A. and Balantekin, A. B. and Heeger, K. and Ramsey-Musolf, M. J. and Bemmerer, D. and Junghans, A. and Bertulani, C. A. and Chen, J.-W. and Costantini, H. and Prati, P. and Couder, M. and Uberseder, E. and Wiescher, M. and Cyburt, R. and Davids, B. and Freedman, S. J. and Gai, M. and Gazit, D. and Gialanella, L. and Imbriani, G. and Greife, U. and Hass, M. and Haxton, W. C. and Itahashi, T. and Kubodera, K. and Langanke, K. and Leitner, D. and Leitner, M. and Vetter, P. and Winslow, L. and Marcucci, L. E. and Motobayashi, T. and Mukhamedzhanov, A. and Tribble, R. E. and Nollett, Kenneth M. and Nunes, F. M. and Park, T.-S. and Parker, P. D. and Schiavilla, R. and Simpson, E. C. and Spitaleri, C. and Strieder, F. and Trautvetter, H.-P. and Suemmerer, K. and Typel, S.},
  journal = {Rev. Mod. Phys.},
  volume = {83},
  issue = {1},
  pages = {195--245},
  numpages = {0},
  year = {2011},
  month = {Apr},
  publisher = {American Physical Society},
  doi = {10.1103/RevModPhys.83.195},
  url = {https://link.aps.org/doi/10.1103/RevModPhys.83.195}
}

@article{Paneru,
    author = "Paneru, S. N. and others",
    title = "{Elastic scattering of He3+He4 with the SONIK scattering chamber}",
    eprint = "2211.14641",
    archivePrefix = "arXiv",
    primaryClass = "nucl-ex",
    doi = "10.1103/PhysRevC.109.015802",
    journal = "Phys. Rev. C",
    volume = "109",
    number = "1",

    pages = "015802",
    year = "2024"
}

@article{Tilley2002,
	author = {D.R. Tilley and C.M. Cheves and J.L. Godwin and G.M. Hale and H.M. Hofmann and J.H. Kelley and C.G. Sheu and H.R. Weller},
	doi = {https://doi.org/10.1016/S0375-9474(02)00597-3},
	issn = {0375-9474},
	journal = {Nuclear Physics A},
	number = {1},
	pages = {3-163},
	title = {Energy levels of light nuclei A=5, 6, 7},
	url = {https://www.sciencedirect.com/science/article/pii/S0375947402005973},
	volume = {708},
	year = {2002}
}

@article{Spiger_1967,
  title = {Scattering of ${\mathrm{He}}^{3}$ by ${\mathrm{He}}^{4}$ and of ${\mathrm{He}}^{4}$ by Tritium},
  author = {Spiger, R. J. and Tombrello, T. A.},
  journal = {Phys. Rev.},
  volume = {163},
  issue = {4},
  pages = {964--984},
  numpages = {0},
  year = {1967},
  month = {Nov},
  publisher = {American Physical Society},
  doi = {10.1103/PhysRev.163.964},
  url = {https://link.aps.org/doi/10.1103/PhysRev.163.964}
}

@article{Odell:2022,
    author = {Odell, Daniel and Brune, Carl R. and Phillips, Daniel R. and deBoer, Richard James and Paneru, Som Nath},
    title = {Performing Bayesian Analyses With AZURE2 Using BRICK: An Application to the 7Be System},
    eprint = {2112.12838},
    archivePrefix = {arXiv},
    primaryClass = {nucl-th},
    doi = {10.3389/fphy.2022.888476},
    journal = {Front. in Phys.},
    volume = {10},
    pages = {888476},
    year = {2022}
}

@article{Critchfield:1949,
  title = {Phase Shifts in Proton-Alpha-Scattering},
  author = {Critchfield, C. L. and Dodder, D. C.},
  journal = {Phys. Rev.},
  volume = {76},
  issue = {5},
  pages = {602--605},
  numpages = {0},
  year = {1949},
  month = {Sep},
  publisher = {American Physical Society},
  doi = {10.1103/PhysRev.76.602},
  url = {https://link.aps.org/doi/10.1103/PhysRev.76.602}
}

@phdthesis{Paneru:thesis,
	address = {Ohio University},
	author = {Paneru, Som N. },
	id = {ohiou1595631779431617},
	school = {Ohio University},
	title = {Elastic Scattering of $^{3}$He+$^{4}$He with SONIK},
	url = {http://rave.ohiolink.edu/etdc/view?acc_num=ohiou1595631779431617},
	year = {2020}
}

@article{Burnelis:2024,
    author = "Burnelis, Andrius and Kejzlar, Vojta and Phillips, Daniel R.",
    title = "{Variational inference of effective range parameters for $^{3}$He\ensuremath{-}$^{4}$He scattering}",
    eprint = "2408.13250",
    archivePrefix = "arXiv",
    primaryClass = "nucl-th",
    doi = "10.1088/1361-6471/ad9296",
    journal = "J. Phys. G",
    volume = "52",
    number = "1",
    pages = "015109",
    year = "2025"
}

@article{Melendez:2017,
    author = "Melendez, J. A. and Wesolowski, S. and Furnstahl, R. J.",
    title = "{Bayesian truncation errors in chiral effective field theory: nucleon-nucleon observables}",
    eprint = "1704.03308",
    archivePrefix = "arXiv",
    primaryClass = "nucl-th",
    doi = "10.1103/PhysRevC.96.024003",
    journal = "Phys. Rev. C",
    volume = "96",
    number = "2",
    pages = "024003",
    year = "2017"
}

@article{Melendez:2019,
  title = {Quantifying correlated truncation errors in effective field theory},
  author = {Melendez, J. A. and Furnstahl, R. J. and Phillips, D. R. and Pratola, M. T. and Wesolowski, S.},
  journal = {Phys. Rev. C},
  volume = {100},
  issue = {4},
  pages = {044001},
  numpages = {22},
  year = {2019},
  month = {Oct},
  publisher = {American Physical Society},
  doi = {10.1103/PhysRevC.100.044001},
  url = {https://link.aps.org/doi/10.1103/PhysRevC.100.044001}
}

@article{Wesolowski:2021,
  title = {Rigorous constraints on three-nucleon forces in chiral effective field theory from fast and accurate calculations of few-body observables},
  author = {Wesolowski, S. and Svensson, I. and Ekstr\"om, A. and Forss\'en, C. and Furnstahl, R. J. and Melendez, J. A. and Phillips, D. R.},
  journal = {Phys. Rev. C},
  volume = {104},
  issue = {6},
  pages = {064001},
  numpages = {14},
  year = {2021},
  month = {Dec},
  publisher = {American Physical Society},
  doi = {10.1103/PhysRevC.104.064001},
  url = {https://link.aps.org/doi/10.1103/PhysRevC.104.064001}
}

@article{Boykin:1972,
	author = {W.R. Boykin and S.D. Baker and D.M. Hardy},
	doi = {https://doi.org/10.1016/0375-9474(72)90732-4},
	issn = {0375-9474},
	journal = {Nuclear Physics A},
	number = {1},
	pages = {241-249},
	title = {Scattering of 3He and 4He from polarized 3He between 4 and 10 MeV},
	url = {https://www.sciencedirect.com/science/article/pii/0375947472907324},
	volume = {195},
	year = {1972}
}

@article{Barnard:1964,
	author = {A.C.L. Barnard and C.M. Jones and G.C. Phillips},
	doi = {https://doi.org/10.1016/0029-5582(64)90235-4},
	issn = {0029-5582},
	journal = {Nuclear Physics},
	pages = {629-640},
	title = {The scattering of He3 by He4},
	url = {https://www.sciencedirect.com/science/article/pii/0029558264902354},
	volume = {50},
	year = {1964}
}

@article{Spiger:1967,
  title = {Scattering of ${\mathrm{He}}^{3}$ by ${\mathrm{He}}^{4}$ and of ${\mathrm{He}}^{4}$ by Tritium},
  author = {Spiger, R. J. and Tombrello, T. A.},
  journal = {Phys. Rev.},
  volume = {163},
  issue = {4},
  pages = {964--984},
  numpages = {0},
  year = {1967},
  month = {Nov},
  publisher = {American Physical Society},
  doi = {10.1103/PhysRev.163.964},
  url = {https://link.aps.org/doi/10.1103/PhysRev.163.964}
}

@article{Higa:2018,
	author = {Higa, Renato and Rupak, Gautam and Vaghani, Akshay},
	journal = {The European Physical Journal A},
	number = {5},
	pages = {89},
	title = {Radiative 3He({\$}{$\backslash$}alpha , {$\backslash$}gamma{\$})7Be reaction in halo effective field theory},
	volume = {54},
	year = {2018}
}

@article{DohetEraly:2016,
	author = {J{\'e}r{\'e}my Dohet-Eraly and Petr Navr{\'a}til and Sofia Quaglioni and Wataru Horiuchi and Guillaume Hupin and Francesco Raimondi},
	doi = {https://doi.org/10.1016/j.physletb.2016.04.021},
	issn = {0370-2693},
	journal = {Physics Letters B},
	pages = {430-436},
	title = {He3($\alpha,\gamma$)Be7 and H3($\alpha,\gamma$)Li7 astrophysical S factors from the no-core shell model with continuum},
	url = {https://www.sciencedirect.com/science/article/pii/S0370269316300946},
	volume = {757},
	year = {2016}
}

@article{Atkinson:2025,
	author = {M.C. Atkinson and K. Kravvaris and S. Quaglioni and P. Navr{\'a}til},
	doi = {https://doi.org/10.1016/j.physletb.2024.139189},
	issn = {0370-2693},
	journal = {Physics Letters B},
	pages = {139189},
	title = {Ab initio calculation of the 3He($\alpha,\gamma$)7Be astrophysical S factor with chiral two- and three-nucleon forces},
	url = {https://www.sciencedirect.com/science/article/pii/S0370269324007470},
	volume = {860},
	year = {2025}
}

@article{Zhang:2020,
    author = "Zhang, Xilin and Nollett, Kenneth M. and Phillips, Daniel R.",
    title = "{$S$-factor and scattering parameters from ${}^3$He + ${}^4$He $\rightarrow {}^7$Be + $\gamma$ data}",
    eprint = "1811.07611",
    archivePrefix = "arXiv",
    primaryClass = "nucl-th",
    reportNumber = "NT@UW-18-15",
    doi = "10.1007/978-3-030-32357-8_30",
    journal = "Springer Proc. Phys.",
    volume = "238",
    pages = "169--173",
    year = "2020"
}

@article{Premarathna:2020,
	author = {Premarathna, Pradeepa and Rupak, Gautam},
	date = {2020/06/12},
	doi = {10.1140/epja/s10050-020-00113-z},
	id = {Premarathna2020},
	isbn = {1434-601X},
	journal = {The European Physical Journal A},
	number = {6},
	pages = {166},
	title = {Bayesian analysis of capture reactions {\$}{\$}{$\backslash$}varvec{\{}\^{}3{\}}{$\backslash$}hbox {\{}He{\}}{$\backslash$}varvec{\{}({$\backslash$}alpha ,{$\backslash$}gamma )\^{}7{\}}{$\backslash$}hbox {\{}Be{\}}{\$}{\$}and {\$}{\$}{$\backslash$}varvec{\{}\^{}3{\}}{$\backslash$}hbox {\{}H{\}}{$\backslash$}varvec{\{}({$\backslash$}alpha ,{$\backslash$}gamma )\^{}7{\}}{$\backslash$}hbox {\{}Li{\}}{\$}{\$}},
	url = {https://doi.org/10.1140/epja/s10050-020-00113-z},
	volume = {56},
	year = {2020}
}

@article{Melendez:2021,
	author = {Melendez, J. A. and Furnstahl, R. J. and Grie{\ss}hammer, H. W. and McGovern, J. A. and Phillips, D. R. and Pratola, M. T.},
	date = {2021/02/27},
	doi = {10.1140/epja/s10050-021-00382-2},
	id = {Melendez2021},
	isbn = {1434-601X},
	journal = {The European Physical Journal A},
	number = {3},
	pages = {81},
	title = {Designing optimal experiments: an application to proton Compton scattering},
	url = {https://doi.org/10.1140/epja/s10050-021-00382-2},
	volume = {57},
	year = {2021}
}

@article{Bedaque:2003,
    author = "Bedaque, P. F. and Hammer, H. W. and van Kolck, U.",
    title = "{Narrow resonances in effective field theory}",
    eprint = "nucl-th/0304007",
    archivePrefix = "arXiv",
    doi = "10.1016/j.physletb.2003.07.049",
    journal = "Phys. Lett. B",
    volume = "569",
    pages = "159--167",
    year = "2003"
}

@article{Bertulani:2002,
    author = "Bertulani, C. A. and Hammer, H. W. and Van Kolck, U.",
    title = "{Effective field theory for halo nuclei}",
    eprint = "nucl-th/0205063",
    archivePrefix = "arXiv",
    doi = "10.1016/S0375-9474(02)01270-8",
    journal = "Nucl. Phys. A",
    volume = "712",
    pages = "37--58",
    year = "2002"
}

@article{Zhang:2017,
    author = "Zhang, Xilin and Nollett, Kenneth M. and Phillips, Daniel R.",
    title = "{Models, measurements, and effective field theory: Proton capture on $^7Be$ at next-to-leading order}",
    eprint = "1708.04017",
    archivePrefix = "arXiv",
    primaryClass = "nucl-th",
    reportNumber = "NT@UW-17-14",
    doi = "10.1103/PhysRevC.98.034616",
    journal = "Phys. Rev. C",
    volume = "98",
    number = "3",
    pages = "034616",
    year = "2018"
}

@article{Wiescher:2025,
    author = "Wiescher, M. and others",
    title = "{Quantum physics of stars}",
    doi = "10.1103/RevModPhys.97.025003",
    journal = "Rev. Mod. Phys.",
    volume = "97",
    number = "2",
    pages = "025003",
    year = "2025"
}

@article{Adelberger:2011,
    author = "Adelberger, E. G. and others",
    title = "{Solar fusion cross sections II: the pp chain and CNO cycles}",
    eprint = "1004.2318",
    archivePrefix = "arXiv",
    primaryClass = "nucl-ex",
    reportNumber = "INT-PUB-10-016, UCB-NPAT-10-001",
    doi = "10.1103/RevModPhys.83.195",
    journal = "Rev. Mod. Phys.",
    volume = "83",
    pages = "195",
    year = "2011"
}

@article{Christy:1961,
title = {$\gamma$ rays from an extranuclear direct capture process},
journal = {Nuclear Physics},
volume = {24},
number = {1},
pages = {89-101},
year = {1961},
issn = {0029-5582},
doi = {https://doi.org/10.1016/0029-5582(61)91019-7},
url = {https://www.sciencedirect.com/science/article/pii/0029558261910197},
author = {R.F. Christy and Ian Duck}
}

@article{Parker:1963,
  title = {${\mathrm{He}}^{3}(\ensuremath{\alpha}, \ensuremath{\gamma}){\mathrm{Be}}^{7}$ Reaction},
  author = {Parker, P. D. and Kavanagh, R. W.},
  journal = {Phys. Rev.},
  volume = {131},
  issue = {6},
  pages = {2578--2582},
  numpages = {0},
  year = {1963},
  month = {Sep},
  publisher = {American Physical Society},
  doi = {10.1103/PhysRev.131.2578},
  url = {https://link.aps.org/doi/10.1103/PhysRev.131.2578}
}

@article{Khadka:2025hef,
    author = "Khadka, Ratna and Gan, Ling and Higa, Renato and Rupak, Gautam",
    title = "{Precision calculation of $^3$He$(\alpha,\alpha)^7$Be for solar physics}",
    eprint = "2509.24931",
    archivePrefix = "arXiv",
    primaryClass = "nucl-th",
    month = "9",
    year = "2025"
}

@article{Angeli:2013epw,
    author = "Angeli, I. and Marinova, K. P.",
    title = "{Table of experimental nuclear ground state charge radii: An update}",
    doi = "10.1016/j.adt.2011.12.006",
    journal = "Atom. Data Nucl. Data Tabl.",
    volume = "99",
    number = "1",
    pages = "69--95",
    year = "2013"
}

@article{Purcell:2010,
title = {Energy levels of light nuclei A=3},
journal = {Nuclear Physics A},
volume = {848},
number = {1},
pages = {1-74},
year = {2010},
issn = {0375-9474},
doi = {https://doi.org/10.1016/j.nuclphysa.2010.08.012},
url = {https://www.sciencedirect.com/science/article/pii/S0375947410006548},
author = {J.E. Purcell and J.H. Kelley and E. Kwan and C.G. Sheu and H.R. Weller}
}
